\pdfoutput=1
\documentclass[prb,twocolumn,showpacs,amsmath,amssymb,floatfix]{revtex4}

\usepackage{graphicx}
\usepackage{dcolumn}
\usepackage{bm}

\begin{document}

\title{Electron density and transport in top-gated graphene nanoribbon devices: First principles Green function algorithms for systems containing large number of atoms}

\author{Denis A. Areshkin and Branislav K. Nikoli\' c}
\affiliation{Department of Physics and Astronomy, University of Delaware, Newark, DE 19716-2570, USA}

\begin{abstract}
The recent fabrication of graphene nanoribbon (GNR) field-effect transistors  poses
a challenge for first-principles modeling of carbon nanoelectronics due to many thousand  atoms present in the device. The state of the art
quantum transport algorithms, based on the nonequilibrium Green function formalism combined with the density functional theory (\mbox{NEGF-DFT}), were originally developed to calculate self-consistent electron density in equilibrium and at finite bias voltage (as a prerequisite to obtain conductance or current-voltage characteristics, respectively) for small molecules attached to metallic electrodes where only a few hundred atoms are typically simulated. Here we introduce combination of two numerically efficient algorithms which make it possible to extend the NEGF-DFT framework to device simulations involving large number of atoms. Our first algorithm offers an alternative to the usual evaluation of the equilibrium part of electron density via numerical contour integration of the retarded Green function in the upper complex  half-plane. It is based on the replacement of the Fermi function $f(E)$ with an analytic function $\tilde{f}(E)$  coinciding with $f(E)$ inside the integration range along the real axis, but decaying exponentially in the upper complex half-plane. Although $\tilde{f}(E)$ has infinite number of poles, whose positions and residues are determined analytically, only a finite number of those poles  have non-negligible residues. We also discuss how this algorithm can be extended to compute the nonequilibrium contribution to electron density, thereby evading  cumbersome real-axis integration (within the bias voltage window) of NEGFs  which is very difficult to converge for systems with large number of atoms while maintaining current conservation. Our second algorithm combines the recursive formulas with the geometrical partitioning of an arbitrary {\em multi-terminal} device into non-uniform segments in order to reduce the computational complexity of the retarded Green function computation by evaluating only its submatrices required for electron density or transmission function. We illustrate fusion of these two algorithms into the \mbox{NEGF-DFT}-type code by computing charge transfer, charge redistribution and conductance in  \mbox{zigzag-GNR$|$variable-width-armchair-GNR$|$zigzag-GNR} two-terminal device covered with a gate electrode made of graphene layer as well. The total number of carbon and edge-passivating hydrogen atoms within the simulated central region of this device is  $\simeq 7000$. Our self-consistent modeling of the gate voltage effect suggests that rather large gate voltage \mbox{$\simeq 3$ eV}  might be required to shift the band gap of the proposed AGNR interconnect and switch the transport from insulating into the regime of a single open conducting channel.
\end{abstract}


\maketitle

\section{Introduction} \label{sec:introduction}

The recent discovery of graphene~\cite{Geim2007}---a single layer of graphite representing first truly two-dimensional crystal~\cite{Neto2009}---has opened new avenues for carbon nanoelectronics.~\cite{Avouris2009,Burghard2009} The limits on continued scaling of present silicon-based electronics are set by the fundamental physical effects~\cite{Keyes2005} (such as quantum tunneling of carriers through the gate insulator and through the body-to-drain junction; dependence of the subthreshold behavior on temperature; and discrete doping effects), the most detrimental being power dissipated in various leakage mechanisms. This is especially dangerous for minimal field-effect transistor (FET)  dimensions and oxide thicknesses. Following the discovery of carbon nanotubes (CNTs), which are rolled up sheets of graphene, the exploration of carbon nanoelectronics over the past decade as a strong contender to aging silicon technology has been centered around semiconducting CNTs as the new type of channel for FET that also makes possible unconventional transistor designs.~\cite{Avouris2009}

Single-wall CNTs bring their unique features into nanoelectronics arena, such as ballistic transport or diffusion with very long mean free paths, high mobility at room temperature due to suppressed electron--acoustic-phonon scattering,  current carrying capacities of the order of $10^9$ A/cm$^2$, and one of the largest known specific stiffness.~\cite{Avouris2009} However, full integration of CNTs into complex high-performance nanoelectronic devices has been thwarted by several unresolved issues, such as: (i) electronic inhomogeneity where random mixture of semiconducting and metallic CNT (due to uncontrolled distribution of diameters and chirality in current synthesis methods) degrade device performance; (ii) difficulty in aligning and patterning through standard lithography methods suitable for high-volume production because of CNTs not being flat; and (iii) extreme sensitivity to minute changes in their local chemical environment.~\cite{Avouris2007}

Graphene shares many of the features of CNT, offering large critical current densities~\cite{Novoselov2004} and intrinsic mobility limit $\simeq 2 \times 10^5$ cm$^2$/Vs  at room temperature being higher than any of the known inorganic semiconductors.~\cite{Chen2008} Such high mobility promises near-ballistic transport and ultrafast switching. Thus, from its inception,~\cite{Novoselov2004} application of graphene in FET devices has been a major experimental endeavor.~\cite{Meric2008,Lin2009}

However, all graphene-FETs fabricated with wide sheets~\cite{Meric2008,Lin2009} have poor ratio of on-state current $I_{\rm on}$ to off-state current $I_{\rm off}$ due to the bulk graphene samples behaving as a zero gap semiconductor. Nevertheless, recent breakthrough fabrication (via chemical derivation,~\cite{Li2008} STM tip drawing~\cite{Tapaszto2008} and CNT unrolling~\cite{Jiao2009,Kosynkin2009}) of sub-10-nm-wide graphene nanoribbons (GNRs), {\em all of which are semiconducting}, has led to the development of \mbox{GNRFETs}~\cite{Wang2008a} with $I_{\rm on}/I_{\rm off}$ ratio up to $10^6$ which is suitable for logic devices.

Moreover, unusual band structure of graphene has generated a plethora of proposals to create devices that have no analog in silicon-based electronics. The new functionality brought by the GNR electronic structure,~\cite{Cresti2008} such as ``valley valves''~\cite{Rycerz2007} or difference in transmission properties of reflectionless $120^\circ$ and highly reflective $60^\circ$ turns made of GNRs with zigzag edges,~\cite{Areshkin2007a} can only be captured by quantum transport analysis. At the same time, equilibrium interatomic charge transfer and chemical doping by different atoms~\cite{Li2008a,Dutta2008,Biel2009} or atomic groups~\cite{Lee2009} that passivate GNR edges require to model explicitly atomistic structure and corresponding charge density within the device. These tasks are beyond the scope of popular tight-binding models~\cite{Rycerz2007,Ouyang2007,Liang2008} (projected onto the basis of single $p_z$-orbital per carbon atom), or even simpler continuous Weyl Hamiltonian describing massless Dirac fermions as low-energy quasiparticles close to the charge neutrality point.~\cite{Neto2009} Furthermore, in the nonequilibrium state driven by the finite bias voltage  one has to compute self-consistently charge redistribution and the corresponding electric potential in order to keep the gauge invariance~\cite{Christen1996} of the \mbox{\em I-V} characteristics~\cite{Areshkin2009} intact.

Finally, virtually every experiment on graphene employees gate electrodes to move the Fermi level away from the charge neutrality point or shift conduction from electron to hole carriers, so that self-consistent computation of the {\em inhomogeneous} charge  distribution~\cite{Fernandez-Rossier2007,Silvestrov2008,Shylau2009} induced by the gate voltage and its highly non-trivial effects on the band structure of GNRs~\cite{Fernandez-Rossier2007,Shylau2009,Guo2007} is necessary to understand device performance (rather than using unrealistic constant shift of the on-site potential to simulate the presence of the gate electrode in the tight-binding models~\cite{Rycerz2007}).

Thus, the prime candidate capable of  handling all of these issues within a unified quantum transport framework~\cite{Stokbro2008,Koentopp2008} is the nonequilibrium Green function (NEGF) formalism~\cite{Haug2007} combined with the density functional theory (DFT) in standard approximation schemes~\cite{Fiolhais2003} (such as LDA, GGA, or B3LYP) for its exchange-correlation potential. The sophisticated algorithms~\cite{Taylor2001,Brandbyge2002,Xue2002,Palacios2002,Ke2004,Pecchia2004,Evers2004,Faleev2005,Rocha2006,Thygesen2008} developed to implement the \mbox{NEGF-DFT} framework over the past decade can be encapsulated by the iterative self-consistent loop:~\cite{Haug2007}
\begin{equation}\label{eq:scloop}
n^{\rm in}({\bf r}) \Rightarrow {\rm DFT} \rightarrow {\bf H}_{\rm KS}[n({\bf r})] \Rightarrow {\rm NEGF} \rightarrow n^{\rm out}({\bf r}).
\end{equation}
The loop starts from the initial input electron density $n^{\rm in}({\bf r})$ $\Rightarrow$ employs some standard DFT code~\cite{Fiolhais2003} (typically in the basis set of finite-range orbitals for the valence electrons which allows for faster numerics and unambiguous partitioning of the system into ``central region'' and the semi-infinite ideal leads) to get the single particle \mbox{Kohn-Sham} Hamiltonian \mbox{${\bf H}_{\rm KS}[n({\bf r})]=-\hbar^2\nabla^2/2m  + V^{\rm eff}({\bf r})$} [\mbox{$V^{\rm eff}({\bf r})=V_H({\bf r})+V_{\rm xc}({\bf r})+V_{\rm ext}({\bf r})$} is the DFT mean-field potential due to other electrons with $V_H({\bf r})$ being the Hartree and $V_{\rm xc}({\bf r})$ the exchange-correlation contribution; $V_{\rm ext}({\bf r})$ is the external potential]  $\Rightarrow$ inversion of ${\bf H}_{\rm KS}[n({\bf r})]$ yields the retarded Green function ${\bf G}^r(E)$ whose integration over energy determines the density matrix via NEGF-based formula:
\begin{widetext}
\begin{equation}\label{eq:rho}
{\bm \rho}  =  -\frac{1}{\pi} \int\limits_{-\infty}^{+\infty} dE \, {\rm Im}\, [{\bf G}^r(E)] f(E-\mu_R)  - \frac{1}{\pi} \int\limits_{-\infty}^{+\infty}dE \,  {\bf G}^r(E)\cdot {\rm Im}\, [{\bm \Sigma}_{L}(E)] \cdot {\bf G}^a(E)  \left[f\left(E-\mu_{L}\right)-f\left(E-\mu_{R}\right)\right] = {\bm \rho}_{\rm eq} + {\bm \rho}_{\rm neq}.
\end{equation}
\end{widetext}
The matrix elements \mbox{$n(\bf r)=\langle {\bf r} | {\bm \rho} | {\bf r} \rangle$} are the new electron density  as the starting point of the next iteration. This procedure is repeated until the convergence criterion $||{\bm \rho}^{\rm out} - {\bm \rho}^{\rm in}|| < \delta$ is reached, where $\delta \ll 1$ is a tolerance parameter.

The representation of the retarded Green function in the local orbital basis requires to compute the inverse matrix
\begin{equation}\label{eq:gr}
{\bf G}^r(E) = [E - {\bf H}_{\rm KS}[n({\bf r})] - {\bm \Sigma}(E)]^{-1}.
\end{equation}
The advanced Green function matrix is defined as ${\bf G}^a(E)=[{\bf G}^r(E)]^\dagger$. The non-Hermitian matrix \mbox{${\bm \Sigma}(E)={\bm \Sigma}_L(E) + {\bm \Sigma}_R(E)$} is the sum of the retarded self-energy matrices introduced by the ``interaction'' with the left [${\bm \Sigma}_{L}(E)$] and the right [${\bm \Sigma}_{R}(E)$] leads. These self-energies determine escape rates of electrons from the central region into the semi-infinite ideal leads, so that an open quantum system can be viewed as being described by the (non-Hermitian) Hamiltonian \mbox{${\bf H}_{\rm open}={\bf H}_{\rm KS}[n({\bf r})] + {\bm \Sigma}(E)$}.

The NEGF post-processing of the converged result of DFT calculations  makes it possible to obtain the current through a  two-terminal device in terms of  the Landauer-type formula~\cite{Haug2007}
\begin{equation}\label{eq:iv}
I(V_{ds})=\frac{2e}{h} \int\limits_{-\infty}^{+\infty} dE\, T(E,V_{ds}) [f(E-\mu_L)-f(E-\mu_R)].
\end{equation}
This integrates the self-consistent transmission function
\begin{equation}\label{eq:transmission}
T(E,V_{ds})= {\rm Tr} \left\{ {\bm \Gamma}_R (E,V_{ds})  {\bf G}^r_{S,1} {\bm \Gamma}_{L}(E,V_{ds})  {\bf G}^a_{1S} \right\},
\end{equation}
for electrons injected at energy $E$ to propagate from the left to the right electrode under the source-drain applied bias voltage $\mu_{L}-\mu_{R}=eV_{ds}$. Here ${\bf G}^r_{S,1}$ is the submatrix of ${\bf G}^r$ whose elements $\langle S | \hat{G}^r | 1 \rangle$ connect orbitals in the first lead supercell (layer denoted as 1) of the extended central region ``sample + portion of the electrodes'' to the last lead supercell (layer denoted as $S$) of the simulated region.

The matrices ${\bm \Gamma}_{L,R}(E)=i[{\bm \Sigma}_{L,R}(E) - {\bm \Sigma}_{L,R}^\dagger(E)]=-2{\rm Im} \, {\bm \Sigma}_{L,R}(E)$ account for the level broadening due to the coupling to the leads.~\cite{Haug2007} A usual assumption about the leads is that the effect of the bias voltage can be taken into account by a rigid shift of their electronic structure, so that ${\bm \Sigma}_{L,R}(E,V_{ds})={\bm \Sigma}_{L,R}(E \mp eV_{ds}/2,0)$ and ${\bm \Gamma}_{L,R}(E,V_{ds})={\bm \Gamma}_{L,R}(E \mp eV_{ds}/2,0)$  are computed in equilibrium and then the shift $\pm eV_{ds}/2$ is applied to their electronic structure to mimic the applied bias. The energy window for the integral in Eq.~(\ref{eq:iv}) is defined by the difference of Fermi functions  $f(E-\mu_L)-f(E-\mu_R)$ of macroscopic reservoirs into which semi-infinite ideal leads terminate. The  formula~(\ref{eq:iv}) is valid only for coherent transport, i.e., assuming absence of dephasing~\cite{Golizadeh-Mojarad2007a} due electron-phonon or electron-electron interactions (beyond those captured by the mean-field treatment~\cite{Thygesen2008,Darancet2007}).

Thus, the most demanding computational task of the \mbox{NEGF-DFT} framework is the self-consistent evaluation of the density matrix ${\bm \rho}$ whose different algorithmic steps have the following~\cite{Stokbro2008} {\em computational complexity}~\cite{Mertens2002} in terms of the number of atoms $N$:~\cite{footnote1} ({\em i}) the computation $n^{\rm in}({\bf r}) \rightarrow V^{\rm eff}({\bf r})$ of the effective potential for  ${\bf H}_{\rm KS}[n({\bf r})]$  has complexity $O(N \log N)$; ({\em ii}) the second step, $V^{\rm eff}({\bf r}) \rightarrow {\bf H}_{\rm KS}[n({\bf r})]$, has complexity $O(N)$; ({\em iii}) usual computation of all elements of the retarded Green function, ${\bf H}_{\rm KS}[n({\bf r})] \rightarrow {\bf G}^r$, requires $O(N^3)$ operations; ({\em iv}) ${\bf G}^r \rightarrow {\bm \rho}$ scales as $O(N)$; and ({\em v}) the final step ${\bm \rho} \rightarrow n^{\rm out}({\bf r})$ also has complexity $O(N)$.   Obviously, the bottleneck is set by the retarded Green function computation. Since \mbox{NEGF-DFT} computational codes~\cite{Taylor2001,Brandbyge2002,Xue2002,Palacios2002,Ke2004,Pecchia2004,Evers2004,Rocha2006} are developed  and tested for small molecules attached to metallic electrodes (where they are successful when coupling between the molecule and the electrodes is strong enough to diminish Coulomb blockade effects~\cite{Koentopp2008}), they typically evaluate all elements of ${\bf G}^r$ by inverting through Eq.~(\ref{eq:gr}) the Hamiltonian of the extended molecule region. Because this has to be done repeatedly through self-consistent loop~(\ref{eq:scloop}), the number of atoms in the extended central region ``molecule + portion of the electrodes'' that can be simulated is limited to few hundreds. This bottleneck also prevents realistic modeling of single or multiple~\cite{Sorensen2009} gate electrodes---instead of an additional layer of atoms covering portion of the central region, one typically employs a uniform electric field in the direction perpendicular to the transport.~\cite{Ke2005,He2008}

A more subtle reason for  the failure of conventionally implemented \mbox{NEGF-DFT} codes when applied to systems containing large number of atoms is the integration in the second term ${\bm \rho}_{\rm neq}$ in Eq.~(\ref{eq:rho}) which  must be performed along the real axis since the integrand is not analytic anywhere in the complex plan.  Although this integration is restricted by the Fermi functions to a segment of the order of the applied bias voltage, a very fine integration grid must be used to capture locations of subband edges (introduced by semi-infinite leads) and broadened molecule orbitals where sharp peaks in the integrand occur. This problem is exacerbated in devices containing large number of atoms where the increasing number of such sharp peaks---due to van Hove singularities in the density of states of the leads  or quasi-bound states present when different contacts throughout the device are not perfectly transparent---can  make it virtually impossible to converge ${\bm \rho}_{\rm neq}$.

The present approach in \mbox{NEGF-DFT} algorithms to deal with this issue is to move the line of integration slightly into the complex plane. However, this effectively adds small imaginary part $i\eta$ to the Hamiltonian ${\bf H}_{\rm open}$  which, therefore, does not conserve current. For example, direct application of this procedure to experimental graphene devices, such as 100 nm long \mbox{GNRFET} of Ref.~\onlinecite{Wang2008a}, would lead to substantial difference between the total current in the left and the right leads. This issue is rarely discussed in the usual \mbox{NEGF-DFT} treatment of transport through relatively short molecules where such violation of current conservation is small.

Some recent attempts to solve it, such as locating the peaks due to quasibound states and patching the non-equilibrium density matrix integral,\cite{Li2007,Joon2007} cannot be applied to large systems with many such peaks. The peaks can be broadened by physical dephasing mechanisms due to electron-electron~\cite{Thygesen2008,Darancet2007} or electron-phonon interactions,~\cite{Golizadeh-Mojarad2007a} but this drastically changes the \mbox{NEGF-DFT} approach by requiring additional and computationally very expensive self-consistent loops to calculate extra self-energy functionals~\cite{Haug2007,Thygesen2008,Darancet2007} due to interactions within the device for which the sparsity of the Hamiltonian matrix \mbox{${\bf H}_{\rm open}$} becomes irrelevant.

Recent efforts~\cite{Sorensen2009,Pecchia2008,Li2007,Joon2007,Kazymyrenko2008,Polizzi2009} to replace some of the algorithms within the NEGF part of the \mbox{NEGF-DFT} scheme, such as unfavorable  computational complexity of brute force matrix inversion~\cite{Pecchia2008} or the real-axis integration~\cite{Li2007,Joon2007} in ${\bm \rho}_{\rm neq}$, have still not led to self-consistent electron density and transport calculations for systems composed of more than about a thousand of atoms.~\cite{Sorensen2009} Here we introduce modified \mbox{NEGF-DFT} scheme which is based on our novel algorithm for the integrations in Eq.~(\ref{eq:rho}) combined with the partitioning the nanostructure of arbitrary shape into slices containing much smaller number of atoms. The Green function matrices of these slices, needed to obtain the electron density within the slice, are computed recursively with much more favorable computational complexity than $O(N^3)$. The number of iteration steps within the self-consistent loop is further reduced, in the case of nanodevices in equilibrium or in quasi-equilibrium situations (e.g., due to by non-zero gate voltage and zero or linear response bias voltage), via modified Broyden mixing scheme for input and output charge density.  We demonstrate the capability of our computational code, termed CANNES (carbon nanoelectronics simulator), to treat multi-terminal structures containing large number of atoms by computing the self-consistent electron density and conductance in the presence of the gate voltage in a graphene nanodevice whose extended central region is composed of $\simeq 7000$ carbon and hydrogen atoms.

The paper is organized as follows. Sec.~\ref{sec:algorithm} elaborates on the ``pole summation'' algorithm for computing  integrals in ${\bm \rho}$. In Sec.~\ref{sec:example} we demonstrate efficiency of our approach by setting up a {\em three-terminal} FET-type device whose source and drain electrodes are made of zigzag graphene nanoribbon (ZGNR) source and drain electrodes while its channel is an armchair GNR (AGNR) of variable width and with sizable energy  gap. The third electrode is gate modeled as a rectangularly-shaped layer of carbon atoms covering the FET channel. The dangling bonds of all graphene layers are terminated by hydrogen atoms. The DFT part of the calculation is carried out using the self-consistent environment-dependent tight-binding model (\mbox{SC-EDTB}) with four orbitals per carbon atom and one orbital per hydrogen atom, which is specifically tailored to simulate eigenvalue spectra, electron densities and Coulomb potential distributions for carbon-hydrogen nanostructures.~\cite{Areshkin2004,Areshkin2005} The combination of ``pole summation'' algorithm with the recursive Green function formulas allows us to compute in Sec.~\ref{sec:example} intricate electric potential distribution  in the space around ZGNR-AGNR-ZGNR FET device, as well as to demonstrate how much voltage has to be applied on the gate electrode to push the device from the off-state due to the gap of AGNR into an on-state enabled by a single transport channel crossing the Fermi level. The computed source-drain conductance as a function of the gate voltage also demonstrates that even at zero gate voltage there is a difference between the non-self-consistent and self-consistent conductance, where the latter takes into account charge transfer between different atomic species or different segments of the device. We conclude in Sec.~\ref{sec:conclusions}.

\section{Self-consistent Algorithms for Electron Density} \label{sec:algorithm}

\begin{figure*}
\includegraphics[scale=0.75,angle=0]{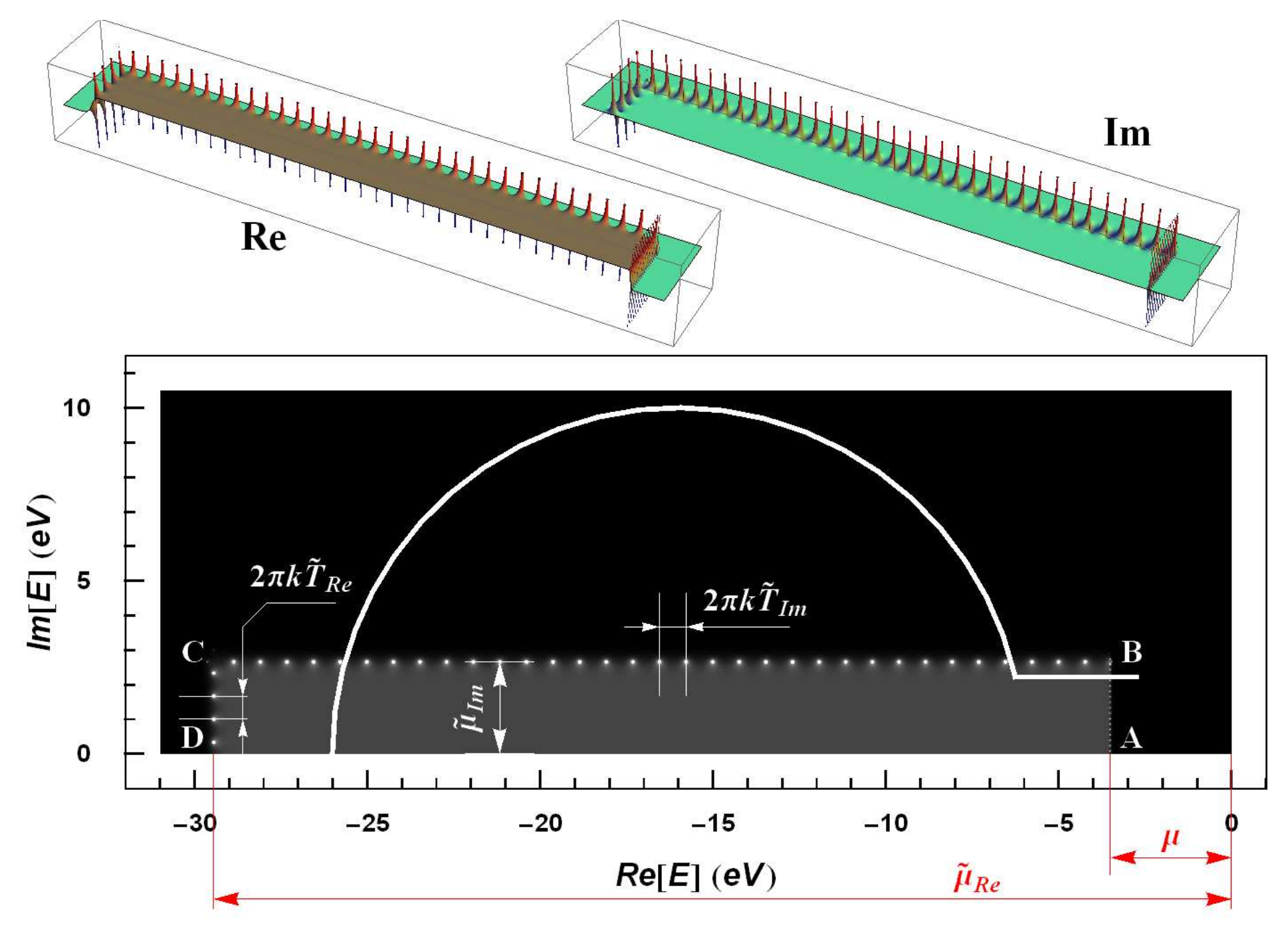}
\caption{(Color online) The density plot of the absolute value of $\tilde{f}(E)$ in the
upper complex half-plane.  Lighter color denotes greater value of
$\mid\tilde{f}\mid$.  Solid black corresponds to zero, while gray color
inside the  dotted rectangle represents unity.  White dots denote the poles with
their size being roughly proportional to the absolute value of the residue.  Poles running
along $AB$, $BC$, and $CD$ edges of the rectangle  correspond  to $z^{(n)}$, $\tilde{z}_{\rm Im}^{(n)}$, and
$\tilde{z}_{\rm Re}^{(n)}$, respectively. Thick white curve denotes the integration contour
traditionally used in \mbox{NEGF-DFT} computational codes.~\cite{Brandbyge2002,Rocha2006}  Top insets are 3D plots of
${\rm Re}\, [\tilde{f}]$ and ${\rm Im}\, [\tilde{f}]$ in the upper complex half-plane.}
\label{fig:PolesDensityPlot}
\end{figure*}

We rewrite the equilibrium contribution to the density matrix~(\ref{eq:rho}):
\begin{eqnarray} \label{subeq:GIntegral}
{\bm \rho}_{\rm eq}(\mu,T) = -\frac{1}{\pi} \int\limits_{E_{\rm min}}^{+\infty} dE\, {\rm Im}\, [{\bf G}^r(E)] f(\mu,T,E),
\end{eqnarray}
in the form which emphasizes its dependence on the chemical potential $\mu$ and temperature $T$, as well as that the
lower limit of integration is the lowest energy at which  ${\rm Im}\, [{\bf G}^r(E_{\rm min})] \neq 0$. As
long as the end-point $E_{\rm min}$ is selected~\cite{Brandbyge2002,Rocha2006} below the bottom of the valence band edge, there are no further poles
in the integrand, and thus the expression is exact. Although this looks obvious, it is important to point out that
if the value $|E_{\rm min}|$ is too small, and there are poles left outside of the contour, the corresponding poles will not be
included in the integration. This causes charge to erroneously  disappear from the system, which typically initiates an avalanche effect,
pushing the poles even further out, and even more  charge is lost, until the system is totally void of electrons. When this occurs, the calculation
will actually converge trivially,  but to a physically incorrect solution.

Since diagonal matrix elements of ${\bf G}^r(E)$ are a rapidly varying function of energy, a direct integration along the real axis would be rather ineffective
since its numerical accuracy is not sufficient to achieve convergence of the self-consistent electron density. Instead, present \mbox{NEGF-DFT}
computational codes~\cite{Taylor2001,Brandbyge2002,Rocha2006} deform the integration contour into the upper complex half-plane ${\rm Im}\, [E]>0$,
where the retarded Green  function is much smoother. This is allowed since ${\bf G}^r(E)$ is analytic in the upper complex
half-plane (all of its poles are slightly displaced below the real axis).

The thick white line in Fig.~\ref{fig:PolesDensityPlot} designates
typically chosen~\cite{Taylor2001,Brandbyge2002,Ke2004,Rocha2006} integration contour. It consists of a semi-circular part $SC$ and a horizontal line $L$ parallel to the real axis
on the right which is positioned to enclose specific number $N_{\rm poles}$ of the Fermi function poles $z^{(n)}$ while ensuring that $SC$ and $L$ are sufficiently far away from the real axis so that the Green function is smooth over both of these two segments [the main variation of the integrand on $L$  comes from the Fermi function $f(E)$ which, therefore, can be used as a weight function in the quadrature~\cite{Brandbyge2002,Rocha2006}]. The final expression for ${\bm \rho}_{\rm eq}$ obtained in this procedure (using the Cauchy residue theorem for the closed contour $SC$ + $L$ + vertical segment from $L$ to the real axis + portion of the real axis) is:
\begin{eqnarray}\label{eq:smeagol}
{\bm \rho}_{\rm eq} & = & -\frac{1}{\pi} {\rm Im}\, \left[ \int\limits_{SC+L} dE\, \, {\bf G}^r(E) f(\mu, T, E) \right. \nonumber \\
&&{} - \left. 2 \pi i k_B T \sum_n^{N_{\rm poles}} {\bf G}^r(z^{(n)})\right],
\end{eqnarray}
where the smoothness of  ${\bf G}^r(E)$ on $SC+L$ contour is exploited to perform the approximate integration in the first term by using a quadrature with a small number of points.~\cite{Brandbyge2002,Rocha2006}

Obviously, it would be advantageous to compute integral in Eq.~(\ref{subeq:GIntegral}) precisely and without worrying about proper selection of parameters for positioning $SC$ and $L$, via a simple summation over a finite set of complex energies akin to the second term of Eq.~(\ref{eq:smeagol}). Here we introduce such an algorithm which makes
possible virtually exact evaluation of ${\bm \rho}_{\rm eq}$ by ``pole summation.'' This algorithm is discussed separately for high temperatures (and/or valence electrons) in Sec.~\ref{subsec:HighTValence} and low temperatures (and/or core electrons) in  Sec.~\ref{subsec:LowTCore}.

\subsection{High temperature and$/$or valence electrons} \label{subsec:HighTValence}

The algorithm for equilibrium density matrix computation discussed in this Section can be used when the inequality
\begin{eqnarray} \label{subeq:EminTokTRatio}
(\mu-E_{\rm min})/k_B T \lesssim 10^{3},
\end{eqnarray}
is satisfied. If Eq.~(\ref{subeq:EminTokTRatio}) is not satisfied, a
slightly more elaborate algorithm described in the next Sec.~\ref{subsec:LowTCore} is needed.  Let us define the desired precision through the
non-negative number $p$, such that the magnitude of the relative error is $\delta \leq
e^{-p}$.  In most cases the machine precision roughly corresponds to
$p=30$, while the practical range of $p$ is usually between 21 and
27.

We start by introducing a function $\tilde{f}$
\begin{eqnarray} \label{subeq:fTildaDefinition}
&&{} \tilde{f}(\mu,\tilde{\mu}_{\rm Re},\tilde{\mu}_{\rm Im},T,\tilde{T}_{\rm Re},\tilde{T}_{\rm Im},E)=f(i\tilde{\mu}_{\rm Im},i\tilde{T}_{\rm Im},E)\times\nonumber\\
&&{} \left(f(\mu,T,E)-f(\tilde{\mu}_{\rm Re},-\tilde{T}_{\rm Re},E)\right),
\end{eqnarray}
where all its arguments except $E$ are limited to real domain and
satisfy the following inequalities ($k_B$ is the Boltzmann constant and $i^2=-1$):
\begin{subequations}\label{subeq:fTildaParameters}
\begin{eqnarray}
&&{} \tilde{T}_{\rm Re}>0,~~
\tilde{T}_{\rm Im}>0, \\
&&{} \tilde{\mu}_{\rm Re} \leqslant E_{\rm min}-p
k_B \tilde{T}_{\rm Re}, \\
&&{} \tilde{\mu}_{\rm Im} \geqslant p k_B
\tilde{T}_{\rm Im}. \label{eq:fTildaParameters_c}
\end{eqnarray}
\end{subequations}
The choice of parameters given by Eq.~(\ref{subeq:fTildaParameters})
guarantees that for real $E \geq E_{\rm min}$ the function $\tilde{f}$
deviates from $f$ by no more than $\delta$.  Therefore the
replacement of $f$ with $\tilde{f}$ in the integrand of
Eq.~(\ref{subeq:GIntegral}) will result in the relative error less
than $\delta$. In the following we assume that $p \geq 21$ so that
$\delta \leq 10^{-9}$.

Thus, for all practical purposes we can state
that (all arguments except $E$ are omitted for brevity)
\begin{eqnarray} \label{subeq:GIntegralApprox}
{\bm \rho}_{\rm eq} = -\frac{1}{\pi} {\rm Im} \, \left[\int\limits_{-\infty}^{+\infty} dE \, {\bf G}^r (E)\tilde{f}(E) \right].
\end{eqnarray}
The poles and residues of the first term in the product on the right-hand side of
Eq.~(\ref{subeq:fTildaDefinition}) are given by
\begin{subequations}\label{eq:PolesAndResiduesIm}
\begin{eqnarray}
&&{} \tilde{z}_{\rm Im}^{(n)}  =  i \tilde{\mu}_{\rm Im}+\pi k_B
\tilde{T}_{\rm Im}(2n+1), \label{subeq:PolesImaginary}
\\\nonumber\\
&&{} {\rm Res} \, \left[f(i\tilde{\mu}_{\rm Im},i\tilde{T}_{\rm Im},z)\right]_{z=\tilde{z}_{\rm Im}^{(n)}}  =  -i k_B \tilde{T}_{\rm Im}. \label{subeq:ResiduesImaginary}
\end{eqnarray}
\end{subequations}
where $n$ is an integer.  Similarly, the poles and residues of $f(\mu,T,E)$ in the second term are
\begin{subequations}\label{eq:PolesAndResiduesPhysical}
\begin{eqnarray}
&&z^{(n)} = \mu+\pi i k_B T(2n+1), \label{subeq:PolesPhysical} \\
&&{\rm Res}\, \left[f(\mu,T,z)\right]_{z=z^{(n)}} = -k_B T,
\label{subeq:ResiduesPhysical}
\end{eqnarray}
\end{subequations}
and for $f(\tilde{\mu}_{\rm Re},-\tilde{T}_{\rm Re},E)$ they are
\begin{subequations}\label{eq:PolesAndResiduesRealFicticious}
\begin{eqnarray}
&&{} \tilde{z}_{\rm Re}^{(n)} = \tilde{\mu}_{\rm Re}+\pi i k_B
\tilde{T}_{\rm Re}(2n+1), \label{subeq:PolesRealFicticious}
\\
&&{} {\rm Res}\, \left[f(\tilde{\mu}_{\rm Re},-\tilde{T}_{\rm Re},z)\right]_{z=\tilde{z}_{\rm Re}^{(n)}}
= k_B \tilde{T}_{\rm Re}. \label{subeq:ResiduesRealFicticious}
\end{eqnarray}
\end{subequations}
\begin{figure*}
\includegraphics[scale=0.75,angle=0]{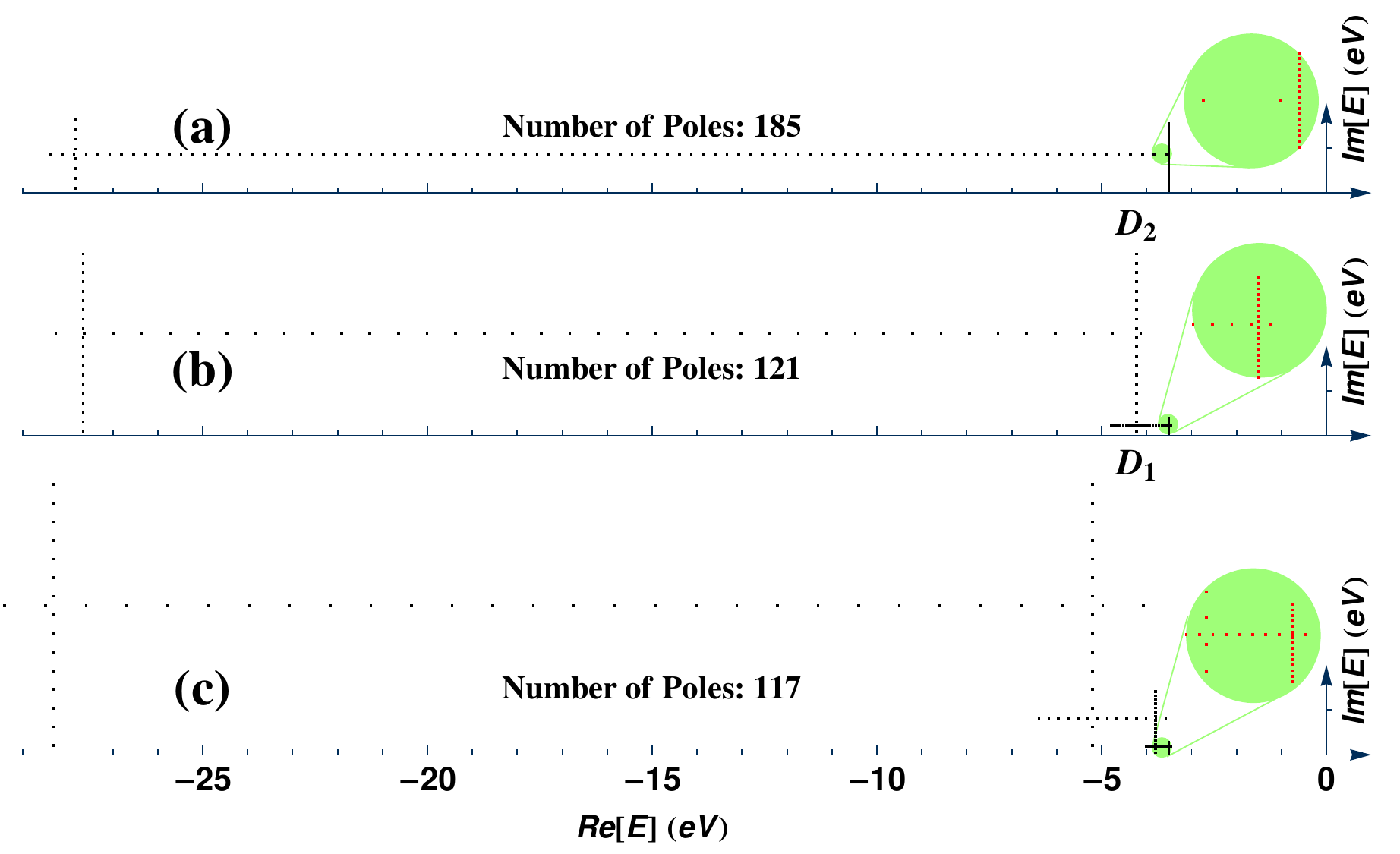}
\caption{(Color online) The poles with non-zero residues for the same system shown
in Fig.~\ref{fig:PolesDensityPlot} but at $k_BT = 0.003$~eV: (a) poles
of $\tilde{f}$; (b) poles of $\tilde{F}^{(2)}$; and (c) poles of $\tilde{F}^{(3)}$.
Circular zoomed-out regions depict dense pole arrangement at
energies close to the chemical potential $\mu$.} \label{fig:MultiStepPolesSetup}
\end{figure*}
Inequalities (\ref{subeq:fTildaParameters}) provide sufficient
freedom to prevent the coincidence of the poles $z^{(j)}$,
$\tilde{z}_{\rm Im}^{(m)}$, and $\tilde{z}_{\rm Re}^{(n)}$ ($\forall$  $j$,
$m$, and~$n$).  Thus, $\tilde{f}$ only has first order poles with residues given by:
\begin{subequations}\label{eq:PolesAndResiduesfTilda}
\begin{eqnarray}
&&{} {\rm Res} \, \left[\tilde{f}(z)\right]_{z=\tilde{z}_{\rm Im}^{(n)}} = -i
k\tilde{T}_{\rm Im}\times\nonumber\\
&&{} \left(f(\mu,T,\tilde{z}_{\rm Im}^{(n)})-f(\tilde{\mu}_{\rm Re},-\tilde{T}_{\rm Re},\tilde{z}_{\rm Im}^{(n)})\right)~,
\label{subeq:ResiduesImaginaryfTilda}
\\
&&{} {\rm Res} \, \left[\tilde{f}(z)\right]_{z=z^{(n)}} = -k_B T
f(i\tilde{\mu}_{\rm Im},i\tilde{T}_{\rm Im},z^{(n)})~,~~~~
\label{subeq:ResiduesPhysicalfTilda}
\\
&&{} {\rm Res}\, \left[\tilde{f}(z)\right]_{z=\tilde{z}_{\rm Re}^{(n)}} = -k_B T
f(i\tilde{\mu}_{\rm Im}, i\tilde{T}_{\rm Im},\tilde{z}_{\rm Re}^{(n)})~.~~~~
\label{subeq:ResiduesRealFicticiousfTilda}
\end{eqnarray}
\end{subequations}
In the upper complex half-plane the residues
(\ref{subeq:ResiduesImaginaryfTilda}) decay exponentially if
${\rm Re}(\tilde{z}_{\rm Im}^{(n)})$ lies outside the interval $
[\tilde{\mu}_{\rm Re},\mu]$, and the residues
(\ref{subeq:ResiduesPhysicalfTilda}),
(\ref{subeq:ResiduesRealFicticiousfTilda}) decay exponentially if
the imaginary component of the poles $z^{(n)}$ or
$\tilde{z}_{\rm Re}^{(n)}$ exceeds $\tilde{\mu}_{\rm Im}$.  Thus, for any
given $p$ only the limited number of poles $\{Z_{j}\}$,
$j\in\{1,N_{\rm pole}\}$ have non-negligible residues.

If one replaces the real axis integration in Eq.~(\ref{subeq:GIntegralApprox}) by the
integration along the semi-circular contour of the sufficiently
large radius in the upper complex half-plane, the contour
contribution to the integral is zero, and the contribution from the
poles is solely from $\{Z_{j}\}$. The integral~(\ref{subeq:GIntegralApprox}) is computed as the sum over all
non-zero residues:
\begin{eqnarray} \label{subeq:GIntegralThroughResidues}
{\bm \rho}_{\rm eq}=-\frac{1}{\pi} {\rm Im}\, \left[\sum_{j=1}^{N_{\rm pole}} {2 \pi i~
{\rm Res} \, \left[\tilde{f}(z)\right]_{z=Z_{j}}{\bf G}^r (Z_{j})}\right],
\end{eqnarray}
where the set $\{Z_{j}\}$ is comprised of  only those $\{\tilde{z}_{\rm Im}^{(n)}\}$,
$\{z^{(n)}\}$, and $\{\tilde{z}_{\rm Re}^{(n)}\}$ poles which satisfy
\begin{subequations}\label{eq:PolesSelection}
\begin{eqnarray}
&&\mid{} f(\mu,T,\tilde{z}_{\rm Im}^{(n)})-f(\tilde{\mu}_{\rm Re},-\tilde{T}_{\rm Re},\tilde{z}_{\rm Im}^{(n)})\mid~\geq
e^{-p}~,~~~ \label{subeq:PolesSelectionImaginary}
\\
&&\mid{} f(i\tilde{\mu}_{\rm Im},i\tilde{T}_{\rm Im},z^{(n)})\mid~\geq e^{-p}~,
\label{subeq:PolesSelectionPhysical}
\\
&&\mid{}
f(i\tilde{\mu}_{\rm Im},i\tilde{T}_{\rm Im},\tilde{z}_{\rm Re}^{(n)})\mid~\geq
e^{-p}~, \label{subeq:PolesSelectionRealFicticious}
\end{eqnarray}
\end{subequations}
respectively, in order to keep the relative error below $e^{-p}$.

For values of $E_{\rm min}$ and $T$ obeying the inequality~(\ref{subeq:EminTokTRatio}) and $21\leq p \leq30$ the number of relevant
poles $N_{\rm pole}$ is moderate.  For example, it is safe to chose $E_{\rm min}=-27$~eV for
valence electrons in a hydro-carbon system (note that this value for $E_{\rm min}$ is measured from the vacuum level). Then, at room temperature the
ratio (\ref{subeq:EminTokTRatio}) is around 700, and for $p=21$ the
minimal number of required poles for parameters satisfying Eq.~(\ref{subeq:fTildaParameters}) equals 76. Decreasing $p$
down to machine precision raises the minimal number of poles to 96.

Figure~\ref{fig:PolesDensityPlot} shows the density plot of
$\tilde{f}$ corresponding to $p=21$ and $E_{\rm min}=-27$~eV used to compute
self-consistent electron within the graphene nanodevice example of Sec.~\ref{sec:example}. The
minimal number of poles $N_{\rm pole}$ is obtained as follows. We consider $\tilde{T}_{\rm Im}$
and $\tilde{T}_{\rm Re}$ as free parameters, and the minimum
allowed  $\tilde{\mu}_{\rm Re}$ and $\tilde{\mu}_{\rm Im}$ are obtained from
equalities in constraints imposed by Eq.~(\ref{subeq:fTildaParameters}).  Then, the
number of poles $z^{(n)}$ is approximately twice the value of
$\tilde{\mu}_{\rm Im}$ divided by the inter-pole distance
\begin{equation}\label{eq:NAB}
N_{AB} = \frac{2 \tilde{\mu}_{\rm Im}}{2 \pi k_B T},
\end{equation}
and the approximate numbers of poles along the lines $CB$ and $DC$ in Fig.~\ref{fig:PolesDensityPlot} are
\begin{subequations}\label{eq:NBCCD}
\begin{eqnarray}
&&{} N_{CB} = \frac{\mu-\tilde{\mu}_{\rm Re}+p k_B \tilde{T}_{\rm Re}+p k_B T}{2
\pi k_B \tilde{T}_{\rm Im}},\label{subeq:NBC}
\\
&&{} N_{DC} = \frac{2 \tilde{\mu}_{\rm Im}}{2 \pi k_B \tilde{T}_{\rm Re}},
\label{subeq:NCD}
\end{eqnarray}
\end{subequations}
respectively. The optimal values of $\tilde{T}_{\rm Im}$ and $\tilde{T}_{\rm Re}$ are
obtained by minimizing $N_{\rm pole}=N_{AB}+N_{CD}+N_{DC}$ in the space of
these two parameters.  A small $\tilde{T}_{\rm Re}$ and
$\tilde{\mu}_{\rm Im}$ adjustment, subject to
constraints~(\ref{subeq:fTildaParameters}), is made afterwards to
place the line $CD$ right in between the two poles on lines $AB$ and
$DC$ (cf. Figs.~\ref{fig:PolesDensityPlot} and \ref{fig:MultiStepPolesSetup}).  This is done to ensure that the
poles are not too close to each other, otherwise a large
numerical errors may occur.

\subsection{Low temperature and$/$or full core simulations} \label{subsec:LowTCore}

The minimum number of poles $N_{\rm pole}$ is scaled by the temperature and the
energy interval $\mu-\tilde{\mu}_{\rm Re}$.  In order to
reduce $N_{\rm pole}$, it is desirable to have as large spacing between the
poles $\tilde{z}_{\rm Im}^{(n)}$ as possible. According to
Eq.~(\ref{eq:fTildaParameters_c}), increasing $\tilde{T}_{\rm Im}$ for the given $p$ means
the increase of $\tilde{\mu}_{\rm Im}$.  The increase of $\tilde{\mu}_{\rm Im}$
in turn increases the length of the segment $AB$, and hence the
number of poles $z^{(n)}$ to be summed. On the other hand, reducing
the number of $z^{(n)}$ (i.e., decreasing $|AB|$=$\tilde{\mu}_{\rm Im}$), will bring the line $CD$ closer to the
real axis, so to prevent deviation of $\tilde{f}$ from
unity on the real axis requires to decrease $\tilde{T}_{\rm Im}$.  The
latter increases the number of poles $\tilde{z}_{\rm Im}^{(n)}$ along
the line $CD$.

The simple solution to this problem is to break the interval between
$\tilde{\mu}_{\rm Re}$ and $\mu$ into several sub-intervals, and apply
the scheme presented in Sec.~\ref{subsec:HighTValence} to each
sub-interval. For example, if the original interval is split into
two sub-intervals, the substitution for $\tilde{f}$ is
\begin{eqnarray} \label{subeq:FTildaExtentionDefinition}
&&{} \tilde{F}^{(2)}(\mu,\tilde{\mu}_{{\rm Re}_{1,2}},\tilde{\mu}_{{\rm Im}_{1,2}},T,\tilde{T}_{{\rm Re}_{1,2}},\tilde{T}_{{\rm Im}_{1,2}},E)=
\nonumber\\
&&{} \tilde{f}(\mu,\tilde{\mu}_{{\rm Re}_{1}},\tilde{\mu}_{{\rm Im}_{1}},T,\tilde{T}_{{\rm Re}_{1}},\tilde{T}_{{\rm Im}_{1}},E)+
\nonumber\\
&&{} \tilde{f}(\tilde{\mu}_{{\rm Re}_{1}},\tilde{\mu}_{{\rm Re}_{2}},\tilde{\mu}_{{\rm Im}_{2}},\tilde{T}_{{\rm Re}_{1}},\tilde{T}_{{\rm Re}_{2}},\tilde{T}_{{\rm Im}_{2}},E),
\end{eqnarray}
where \mbox{$T<\tilde{T}_{{\rm Re}_{1}}<\tilde{T}_{{\rm Re}_{2}}$};
\mbox{$\tilde{\mu}_{{\rm Re}_{2}}<\tilde{\mu}_{{\rm Re}_{1}}<\mu$};
and \mbox{$\tilde{\mu}_{{\rm Im}_{1}}<\tilde{\mu}_{{\rm Im}_{2}}$}. The parameters
$\tilde{\mu}_{{\rm Re}_{1,2}}$, $\tilde{\mu}_{{\rm Im}_{1,2}}$,
$\tilde{T}_{{\rm Re}_{1,2}}$, and $\tilde{T}_{{\rm Im}_{1,2}}$ ensure the
required precision by satisfying the constraints similar to
Eq.~(\ref{subeq:fTildaParameters}):
\begin{subequations}
\begin{eqnarray} \label{subeq:FTildaExtentionParameters}
&&{} \tilde{\mu}_{{\rm Re}_{2}} \leqslant E_{\rm min}-p k_B \tilde{T}_{{\rm Re}_{2}}, \\
&&{} \tilde{\mu}_{{\rm Im}_{1}} \geqslant p k_B
\tilde{T}_{{\rm Im}_{1}},~~\tilde{\mu}_{{\rm Im}_{2}} \geqslant p k_B
\tilde{T}_{{\rm Im}_{2}}.
\end{eqnarray}
\end{subequations}
Figure~\ref{fig:MultiStepPolesSetup}(b) illustrates these concepts.
Poles forming the left (smaller) and the right (bigger) rectangles
are associated respectively with the first and the second term in
Eq.~(\ref{subeq:FTildaExtentionDefinition}). The poles running along
the line $D_{1}D_{2}$ are the same for the first and second term in
Eq.~(\ref{subeq:FTildaExtentionDefinition}).

The minimization of
the total number of poles $N_{\rm pole}$ is performed analogously
to Eqs.~(\ref{eq:NAB}) and (\ref{eq:NBCCD}). For $\tilde{F}^{(2)}$ the
optimization parameters are $\tilde{T}_{{\rm Re}_{1}}$,
$\tilde{T}_{{\rm Re}_{2}}$, $\tilde{T}_{{\rm Im}_{1}}$, and
$\tilde{T}_{{\rm Im}_{1}}$.  The starting point for the conjugate
gradient minimization is $\tilde{T}_{{\rm Re}_{1}} = 10\times T$ and
$\tilde{\mu}_{{\rm Im}_{2}} = 10\times \tilde{\mu}_{{\rm Im}_{1}}$, so that
the optimized parameters fit this order of magnitude relationship.
Indeed, the size of the integration intervals in
Fig.~\ref{fig:MultiStepPolesSetup}(b) and ~\ref{fig:MultiStepPolesSetup}(c) increases by an order of
magnitude from right to left.  For this reason $N_{\rm pole}$ grows
logarithmically with increasing ratio  $(\mu-E_{\rm min})/k_B T$. That is,
depending on $p$, approximately 30 to 40 extra poles are required for each
decade of this ratio increase (i.e., per order of magnitude in temperature reduction).

\subsection{Approximate real axis integration of non-analytic functions} \label{subsec:NonAnalytic}

The concepts presented in Sec.~\ref{subsec:HighTValence}
allow for efficient and exact evaluation of the ${\bf G}^r(E)$ moments in the
interval bounded by two Fermi functions. This property can be used for
systematic approximation of ${\bf G}^a(E)$ with the function
$\tilde{\bf G}^{a}(E)$ such that $\tilde{\bf G}^{a}(E) \approx {\bf G}^a(E)$ on
the real axis, and which is analytic in the upper complex half-plane. This
approximation can be used to transform the non-analytic integrands to
analytic functions.

An obvious applications of this idea to NEGF-DFT framework would be the computation of nonequilibrium
contribution ${\bm \rho}_{\rm neq}$ to the density matrix in Eq.~(\ref{eq:rho}). Because the functions
${\bf G}^{r}(E)$ and ${\bf G}^{a}(E)$ in the integrand of ${\bm \rho}_{\rm neq}$ are non-analytic
below and above the real axis, respectively, the integrand  is non-analytic function in the entire
complex energy plane. Thus, no integration contour deformation akin to Fig.~\ref{fig:PolesDensityPlot} can be
exploited to avoid direct integration along the real axis to obtain ${\bm \rho}_{\rm neq}$. On the other hand,
such direct integration along the real axis is computationally expensive due to the need for very fine integration grids.~\cite{Li2007,Joon2007}
As discussed in Sec.~\ref{sec:introduction}, integration may not even converge  when the integrand becomes too spiky with numerous closely spaced  sharp peaks for devices
containing  large number of atoms.

Let us divide the interval $\left[\mu_R,\mu_L\right]$ into
$M$ subintervals of equal size $\Delta\mu$
\begin{eqnarray}
\label{subeq:IntervalSubdivision}
\mu_{0}=\mu_{R},~~\mu_{M}=\mu_{L},~~\mu_{m}=\mu_{R}+m\Delta\mu,
\end{eqnarray}
where we assume for simplicity that $\Delta\mu=2 k_B T$.
Then ${\bm \rho}_{\rm neq}$ in Eq.~(\ref{eq:rho}) can be rewritten as
\begin{eqnarray}
\label{subeq:NonEquilibriumIntegralSum}
{\bm \rho}_{\rm neq} & = & \sum_{m=1}^{M} \int\limits_{-\infty}^{+\infty} dE\, {\bf G}^{r}(E) \cdot
{\rm Im}\, \left[{\bm \Sigma}(E) \right] \cdot
{\bf G}^a (E) \times \nonumber \\
&&{} \left[f(\mu_{m},T,E)-f(\mu_{m-1},T,E)\right].
\end{eqnarray}
For each interval $\left[\mu_{m-1},\mu_{m}\right]$ in the sum
(\ref{subeq:NonEquilibriumIntegralSum}) we approximate ${\bf G}^{r}(E)$ by
the power expansion with respect to the deviation from the center of
the interval \mbox{$\xi_{m}=(\mu_{m-1}+\mu_{m})/2$}
\begin{eqnarray}
\label{subeq:GExpansion} {\bf G}^r_{m}(E)\approx\tilde{\bf G}^{r}_{m}(E)=
\sum_{\kappa=0}^{K}{{\bf g}_{m}^{(\kappa)}\times(E-\xi_{m})^{\kappa}},
\end{eqnarray}
where ${\bf g}_{m}^{(\kappa)}$ are constant matrices.  We require that the
moments ${\bf M}_{m}^{(d)}$ up to order $D$ for $\tilde{\bf G}^{r}_{m}$ and ${\bf G}^r_{m}$ coincide
\begin{eqnarray} \label{subeq:EqualMoments}
&&{} {\bf M}_{m}^{(d)} \equiv \int\limits_{-\infty}^{+\infty} dE\, {\bf G}^{r}(E) (E-\xi_{m})^{d} \times \nonumber \\
&&{} \left[f(\mu_{m},T,E)-f(\mu_{m-1},T,E)\right] = \nonumber \\
&&{} \int\limits_{-\infty}^{+\infty} dE\, \sum_{\kappa=0}^{K} {\bf g}_{m}^{(\kappa)}\times(E-\xi_{m})^{\kappa+d} \times \nonumber\\
&&{} \left[f(\mu_{m},T,E)-f(\mu_{m-1},T,E)\right],
\end{eqnarray}
where $d\subset[0,D]$.

\begin{figure}
\includegraphics[scale=0.55,angle=0]{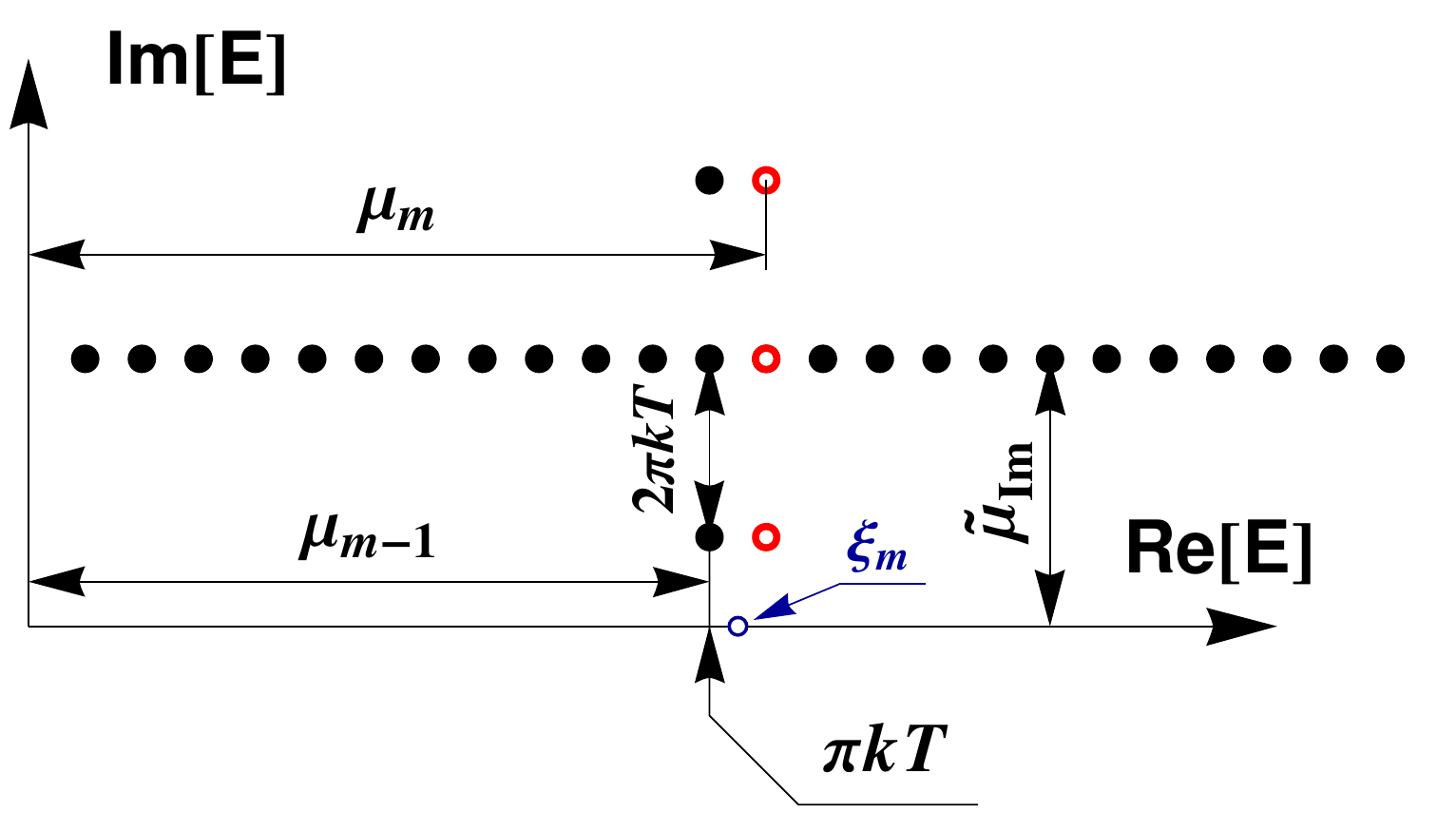}
\caption{(Color online) Poles of the function
$\tilde{f}(\mu_{m},\mu_{m-1},\tilde{\mu}_{\rm Im},T,T,\tilde{T}_{\rm Im},z)$
used to evaluate the integral in Eq.~(\ref{subeq:ExactMoments}).  For a chosen
precision set by $p=23$, the contribution from 28 poles has to be summed.  Three poles of
$f(\mu_{m},T,z)$ are marked with the red empty circles.  The values
of the retarded Green function at most of the poles shown are reused to
compute matrices ${\bf M}^{(d)}_{n}$ in Eq.~(\ref{subeq:EqualMoments}) for $n \neq m$, so that the
average number of Green functions to be computed per interval equals 3.}
\label{fig:MomentRepresentation}
\end{figure}

The first integral in Eq.~(\ref{subeq:EqualMoments}) can be computed accurately as
\begin{eqnarray}
\label{subeq:ExactMoments}
&&{} \int\limits_{-\infty}^{+\infty} dE\, {\bf G}^{r}(E)(E-\xi_{m})^{d} \times \nonumber\\
&&{} \left[f(\mu_{m},T,E)-f(\mu_{m-1},T,E)\right] = \nonumber \\
&&{} \int\limits_{-\infty}^{+\infty}dE\, {\bf G}^{r}(E)(E-\xi_{m})^{d} \times \nonumber\\
&&{} \tilde{f}(\mu_{m},\mu_{m-1},\tilde{\mu}_{\rm Im},T,T,\tilde{T}_{\rm Im},E).
\end{eqnarray}
Figure~\ref{fig:MomentRepresentation} shows the poles of
$\tilde{f}$ from Eq.~(\ref{subeq:ExactMoments}) for the case $p=23$,
$\tilde{\mu}_{\rm Im}=3 \pi k_B T$, and $\tilde{T}_{\rm Im}=T/\pi$. Even
though the number of poles to be summed per every moment equals 28,
the number of points per integration interval $\Delta\mu$ at which
${\bf G}^{r}(E)$ needs to be calculated is 3 because the values of ${\bf G}^{r}(E)$
at different poles are reused in computation of the moments at
different intervals.  Thus, ${\bf M}_{m}^{(d)}$ is computed similarly to
Eq.~(\ref{subeq:GIntegralThroughResidues}), with the only difference being
that ${\bf G}^{r}(Z_{j})$ is now replaced by
${\bf G}^{r}(Z_{j})(Z_{j}-\xi_{m})^{d}$.

Because matrices ${\bf g}_{m}^{(\kappa)}$ do not depend on energy, the
integrals in the second term of Eq.~(\ref{subeq:EqualMoments})
\begin{eqnarray}\label{subeq:UpsilonDefinition}
\Upsilon_{\kappa} & \equiv & \int\limits_{-\infty}^{+\infty}dE \, {{(E-\xi_{m})^{\kappa}}} \times \nonumber \\
&&{} \left[f(\mu_{m},T,E)-f(\mu_{m-1},T,E)\right],
\end{eqnarray}
can be computed analytically. Here we provide example solution of
this problem for $D=2$ (the solutions for $D>2$ are similar to this).
The integrals $\Upsilon_{\kappa}$ are non-zero when $\kappa$ is even integer.
For example, assuming $\Delta \mu = 2 k_B T$ they are
\begin{eqnarray}
\label{subeq:UpsilonDEqual2} &&{} \Upsilon_{0}=2 k_B T,~~\Upsilon_{2}=\frac{2}{3}(k_B
T)^{3}\left(1+\pi^{2}\right),\nonumber\\
&&{} \Upsilon_{4}=\frac{2}{15}(k_B
T)^{5}\left(3+10\pi^{2}+7\pi^{4}\right).
\end{eqnarray}
Then, to satisfy Eq.~(\ref{subeq:EqualMoments}) for $d=0,1,2$,
matrices ${\bf g}_{m}^{(\kappa)}$ should be chosen as
\begin{subequations}
\begin{eqnarray}\label{subeq:gmDEqual2}
{\bf g}_{m}^{(0)} & = &\frac{M_{m}^{(2)}\Upsilon_{2}-M_{m}^{(0)}\Upsilon_{4}}{\Upsilon_{2}^{2}-\Upsilon_{0}\Upsilon_{4}},  \\
\mbox{} {\bf g}_{m}^{(1)} & = & \frac{M_{m}^{(1)}}{\Upsilon_{2}},  \\
\mbox{} {\bf g}_{m}^{(2)} & = & \frac{M_{m}^{(2)}\Upsilon_{0}-M_{m}^{(0)}\Upsilon_{2}}{-\Upsilon_{2}^{2}-\Upsilon_{0}\Upsilon_{4}}.
\end{eqnarray}
\end{subequations}
The analytic continuation of $\tilde{\bf G}^{a}(E)$ into the upper complex
half-plane is simply
\begin{eqnarray}
\label{subeq:GTildaConjugate} &&{} \tilde{\bf G}^{a}_{m}(z)=
\sum_{\kappa=0}^{2} \left[{\bf g}_{m}^{(\kappa)}\right]^\dagger \times (z-\xi_{m})^{\kappa}.
\end{eqnarray}
Then Eq.~(\ref{subeq:NonEquilibriumIntegralSum}) becomes
\begin{eqnarray}
\label{subeq:NonEquilibriumIntegralSumApprox}
&&{} {\bm \rho}_{\rm neq}=\frac{1}{2i} \sum_{m=1}^{M} {\left({\bm \Omega}_{m} - {\bm \Omega}_{m}^{\dag}\right)},
\end{eqnarray}
where
\begin{eqnarray}
\label{subeq:NonEquilibriumIntegralSumApproxOmega}
{\bm \Omega}_{m} & = & \int\limits_{-\infty}^{+\infty}dE\, {{\bf G}^{r}(E) \cdot {\bm \Sigma}(E) \cdot \tilde{\bf G}^{a}(E)} \times
\nonumber \\
&&{} \left[ f(\mu_{m},T,E)-f(\mu_{m-1},T,E) \right].
\end{eqnarray}
The integrand in Eq.~(\ref{subeq:NonEquilibriumIntegralSumApproxOmega}) is now
analytic in the upper-half plane and can be evaluated through our  ``pole
summation'' algorithm discussed in Sections~\ref{subsec:HighTValence} and ~\ref{subsec:LowTCore}.

The algorithm presented in this Section is actually more computationally expensive than
the usually implemented~\cite{Brandbyge2002,Ke2004,Rocha2006} real axis integration to get
${\bm \rho}_{\rm neq}$ since for every interval one needs to
compute the retarded Green function at three different points instead of
one, as shown in Fig.~\ref{fig:MomentRepresentation}.  Nonetheless, the benefit of this
approach is in systematic approximation by exact match of the Green  function moments which
can evade insufficiently fine integration grid or, most importantly, uncontrolled usage~\cite{Brandbyge2002,Faleev2005,Rocha2006} of the
real-axis infinitesimal ${\bf H}_{\rm open}  + i\eta$ that leads to serious current non-conservation in long devices beyond molecular
electronics scale. For example, a very large system poorly coupled to its contacts may have several sharp peaks within 10 meV interval.
None of the adaptive real-axis integration methods~\cite{Li2007,Joon2007} can properly account for these peaks if the integration
step equals  10 meV, while the moments-matching algorithm has capability to capture the contribution from these peaks to the integral.

\section{Example: First-principles modeling of top-gated GNR-based nanoelectronic devices} \label{sec:example}

From the very outset, the discovery of graphene has been intimately connected to attempts to fabricate carbon-based planar FETs.~\cite{Novoselov2004}
Since FETs produced using micron-size graphene sheets as channels have poor $I_{\rm on}/I_{\rm off} \lesssim 10$ ratio, the pursuit of
FETs suitable for digital electronics applications has shifted toward fabrication of GNRs with large band gaps~\cite{Li2008} $\simeq 0.4$~eV. Their
band gap can be engineered by transverse quantum confinement effects in the case of AGNR (where the gap is additionally affected by the increased hopping integral
between the $p_z$-orbitals on carbon atoms around the armchair edge caused by slight changes in atomic bonding length in the presence of edge passivating
hydrogen~\cite{Son2006a}) or by staggered sublattice potential arising due to non-zero spin polarization around zigzag edges of
ZGNR.~\cite{Son2006a,Pisani2007,Yazyev2008,Huang2008,Areshkin2009}

The very recent experiments\cite{Tapaszto2008,Wang2008a,Li2008,Jiao2009,Kosynkin2009} have demonstrated that all sub-10-nm-wide GNRs are semiconducting.
Since band gaps due to edge magnetic ordering in ZGNR are easily destroyed at room temperature,~\cite{Yazyev2008} by finite current under nonequilibrium bias
voltage conditions,~\cite{Areshkin2009} or by impurities and vacancies along the edge,~\cite{Huang2008} we assume that AGNRs are essential ingredient
to introduce sizable band gap in graphene nanodevices operating at room temperature, as confirmed also by recent tunneling spectroscopy.~\cite{Ritter2009}

\begin{figure}
\includegraphics[scale=0.45,angle=0]{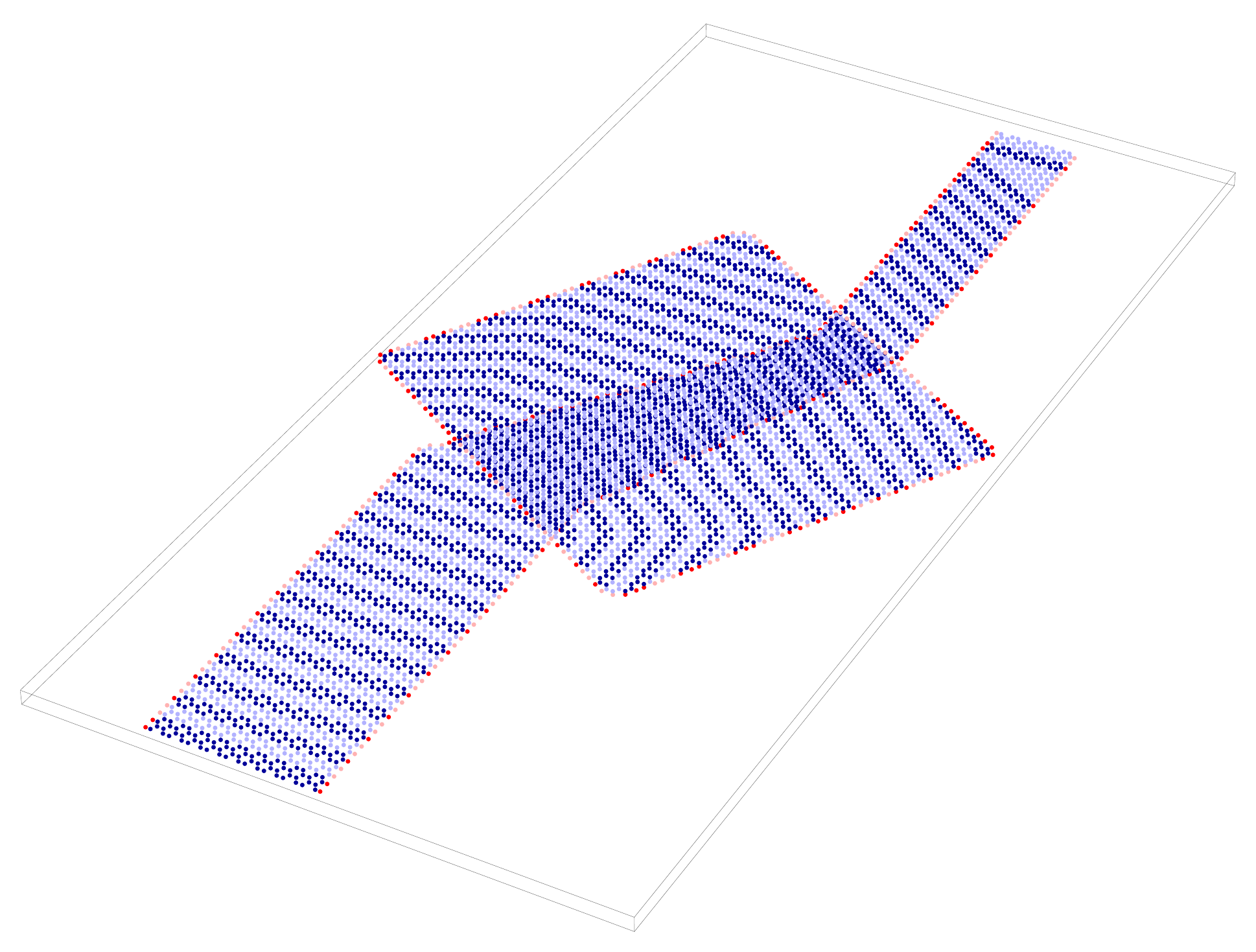}
\caption{(Color online) Graphical depiction of the atomic structure of simulated nanodevice composed of two
narrow graphene layers. The lower graphene layer contains two unidirectional ZGNRs
of different width, which act as the source and drain metallic electrodes, sandwiching semiconducting AGNR of variable width as the FET channel. The top graphene
layer plays the role of a gate electrode, covering all of the AGNR channel region, and has the shape of a rectangle that is sufficiently large to have negligible band gap.
The interlayer distance is 3.35~{\AA}, which corresponds to the interlayer spacing in graphite.~\cite{Girifalco1956} The hydrogen atoms (red dots) passivate edges of both
layers, whose internal carbon atoms (blue) form defect free finite-size honeycomb lattice. Dark and light colored
transverse segments, which have variable shape as one moves from the source to the drain electrode, are used to mark odd and even
slices of the partitioned system. Each slice $i=1$,\ldots,$S$ is described by the Hamiltonian matrix ${\bf H}_{i,i}$, all of which are
stored in computer memory together with matrices ${\bf H}_{i,i+1}$ describing the coupling between adjacent slices $i$ and $i+1$.}
\label{fig:SystemConfiguration}
\end{figure}

The fabricated \mbox{GNRFETs} thus far have utilized metallic source and drain electrodes where Schottky barrier (SB) is introduced at the contact between
metallic electrode (typically Pd with high  work function) and GNR, so that the current is modulated  by carrier tunneling
probability through SB at contacts. On the other hand, planar structure of graphene is envisaged to make possible \mbox{all-graphene} electronic
circuits patterned from either a single graphene plane or multiple planes separated by layers of insulating material.~\cite{Areshkin2007a}

Any \mbox{\em all-graphene} circuit concept will require both active FETs and  passive elements for wiring individual circuit elements. Although
ZGNR can be expected to be metallic at room temperature, the wiring based on them is nontrivial issue because only few specific ZGNR patterns have close
to ideal conductance and can transmit electron flux without losses.~\cite{Areshkin2007a} Furthermore, at finite bias voltage ZGNRs can open a band gap
if they are mirror symmetric with respect to the midplane between the two zigzag edges.~\cite{Li2008a}

\subsection{Three-terminal device setup} \label{subsec:SystemSetup}

Our FET-type device setup, based on the combination of ZGNR source and drain metallic electrodes and semiconducting AGNR channel in between them,
is shown in Fig.~\ref{fig:SystemConfiguration} The source and drain  have different widths and are modeled
as semi-infinite ideal ZGNRs leads. The size of the AGNR band gap is an oscillating function of the ribbon width.  The width variation causing
AGNR to switch between small and large gaps equals to just a single \mbox{C-C} bond length, which was found to greatly affect the transfer
characteristics (current $I$ vs. gate voltage $V_{gs}$ at fixed source-drain bias $V_{ds}$) in the recent study~\cite{Liang2008} of  several FET concepts
with AGNR channel. Because cutting graphene with atomic precision in order to obtain uniform device performance is currently not an option, the variable width AGNR seems to be the
simplest realistic path toward making a short semiconducting fragment. Above the semiconducting ``active region'' we place a graphene rectangle,
which is assumed to have no electrical contact with the ZGNR-AGNR-ZGNR structure below it.  This may be achieved by placing boron-nitride insulating layer in between.

We note that the recent analysis~\cite{Liang2008}  (using NEGF for simple $p_z$-orbital tight-binding model, which is self-consistently coupled to a three-dimensional
Poisson solver for treating the electrostatics) of dual-gate Schottky barrier \mbox{GNRFETs}, with uniform width AGNR channels and several different
types of graphene- or non-graphene-based source and drain electrodes, has singled out ZGNR-AGNR-ZGNR device concept as an optimal one with high
enough $I_{\rm on}/I_{\rm off}$ ratio  and advantageous features  of ZGNR metallic contacts.

The usage of wide graphene sheets as the channel of FET is conceptually difficult because depending on the position of the Fermi level graphene possesses either
electron or hole conductivity making it impossible to produce regions depleted of mobile charge carriers. At the same time, the concept of GNR devices allows to
build both normally-OFF and normally-ON transistors based solely on the device geometries.~\cite{Rycerz2007,Areshkin2007a} One of the main benefits of graphene in nanoelectronics
is its one-atom-thickness which leads to very low parasitic capacitance, and therefore allows terahertz cut-off frequencies for \mbox{all-graphene} devices and circuits.  So far both the experiments~\cite{Wang2008a}  and quantum transport simulations~\cite{Ouyang2007,Liang2008} have been focused on \mbox{GNRFETs}   whose channel is long and narrow semiconducting GNR  attached to metallic source and drain (such as Pd) contacts while being controlled by metallic  top-gate shifting the band gap. Although such transistors play an important role in studying GNR properties, they compromise the main purpose of nanoelectronic devices---the speed. The parasitic gate-substrate or gate-source (drain) capacitances~\cite{Fernandez-Rossier2007,Shylau2009,Guo2007} for such hybrid metal-graphene structure are orders of magnitude higher than capacitance of the channel, and thus substantially decrease the transistor speed. Exploring \mbox{all-graphene} nanoelectronic devices to reach the optimal speed limit is one of the primary motivations for the design concept shown in Fig.~\ref{fig:SystemConfiguration}.

\subsection{System partitioning and the recursive Green function algorithm}\label{sec:recursive}

The retarded Green function matrix ${\bf G}^r(E)$, as the central NEGF quantity in phase-coherent transport regime which yields electron density through Eq.~(\ref{eq:rho}) and current via  Eq.~(\ref{eq:iv}),  can be computed by direct matrix inversion in Eq.~(\ref{eq:gr}). However, the computational complexity $O(N^3)$ of this operation makes it virtually impossible for present \mbox{NEGF-DFT} codes (which typically perform this brute force operation) to be applied to systems containing large number of atoms.~\cite{Stokbro2008} Thus, first-principles simulation of transport in large systems can be accomplished only if relevant elements of  ${\bf G}^r(E)$ can be obtained via algorithms that scale linearly with increasing length of assumed quasi-one-dimensional (Q1D) device geometry.~\cite{Stokbro2008}

In fact, since only a much smaller submatrix of ${\bf G}^r(E)$  determines transport properties given by Eq.~(\ref{eq:iv}), the recursive Green function algorithms~\cite{Ferry1999} (in serial or parallel implementation~\cite{Drouvelis2006}) have commonly been used to compute the submatrix ${\bf G}_{S,1}^r$ and obtain the transmission properties of  mesoscopic devices.~\cite{Ferry1999} They are  based on using the Dyson equation, ${\bf G}^r_C = {\bf G}_0^r + {\bf G}_0^r {\bf V} {\bf G}^r_C$, to build the Green-function slice by slice, so that  the dimensions of the matrices that have to be inverted are strongly reduced (${\bf G}_0^r$ is the Green function of some region of the device  with one of the leads attached, ${\bf V}$
is the hopping matrix between that region and adjacent slice, and ${\bf G}^r_C$ is the Green function of the coupled system lead + region + slice).

This type of algorithms have also been extended~\cite{Cresti2003,Metalidis2005,Lassl2007,Pecchia2008,Nikoli'c2010} to obtain other submatrices of  ${\bf G}^r$ needed to compute  local quantities within the simulated region, such as ${\bf G}^r_{i,i}$ or  ${\bf G}^r_{i,i+1}$ which define the electron density within slice $i$ or spatial profile of local currents between slices $i$ and $i+1$, respectively. Although it is often considered~\cite{Kazymyrenko2008} that standard or extended recursive Green function algorithms can be applied only to Q1D two-terminal devices, some alternative approaches which invert smaller matrices than the full device Hamiltonian ${\bf H}_{\rm open}$ to build the Green function of multi-terminal nanostructures of arbitrary geometrical shape have also been introduced recently.~\cite{Kazymyrenko2008,Nikoli'c2010}

The key issue for a successful inclusion of the recursive Green function formulas into \mbox{NEGF-DFT} codes is not the specific set of equations, which is very similar in different approaches, but the ability to make a consistent partition of a system of arbitrary shape and with many attached  electrodes into slices described by much smaller matrices ${\bf H}_{i,i}$. The full Hamiltonian matrix can then be written as
\begin{widetext}
\begin{equation}\label{eq:hmatrix}
{\bf H}_{\rm KS}=\left(
\begin{array}{cccccc}
{\bf H}_{1,1} &  {\bf H}_{1,2} & 0 & 0 & \cdots & 0 \\
{\bf H}_{1,2}^\dagger & \ddots & \cdots & \cdots &  \cdots & 0 \\
\vdots & {\bf H}_{i-1,i-1} & {\bf H}_{i-1,i} &  0 & \cdots & \vdots \\
\vdots & {\bf H}_{i-1,i}^\dagger  & {\bf H}_{i,i} & {\bf H}_{i,i+1} & \cdots & \vdots \\
\vdots & 0 &  {\bf H}_{i,i+1}^\dagger & {\bf H}_{i+1,i+1} & \cdots & \vdots \\
0 & \cdots & \cdots & \cdots & \ddots & {\bf H}_{S-1,S} \\
0 & 0 & 0 & \cdots & {\bf H}_{S-1,S}^\dagger & {\bf H}_{S,S} \\
\end{array}\right).
\end{equation}
\end{widetext}
since due to the finite range of basis functions in the transport direction the size of the slices can always be chosen so large that only neighboring ones are coupled through each other via the hopping matrices ${\bf H}_{i,i+1}$.

An example of the solution to this primarily {\em geometrical problem} is illustrated using the device setup in Fig.~\ref{fig:SystemConfiguration}. Our algorithm here starts from the bitmap image of the device $\rightarrow$ converts the image into a finite-size honeycomb lattice $\rightarrow$ then attempts to partition the device within a loop until consistent set of slices is achieved across the whole device. The final result---a set of slices of non-uniform shape (in contrast to typical columns of sites orthogonal to the axis of the device when recursive algorithm is applied to two-terminal Q1D devices of simple shape~\cite{Metalidis2005,Lassl2007})---is shown in Fig.~\ref{fig:SystemConfiguration} as dark and light colored segments of the honeycomb lattice.

Each slice is described by a  matrix ${\bf H}_{i,i}$ containing the interactions between atoms within the layer $i$ ($i = 1$,\ldots,$S$). The size of the matrix ${\bf H}_{i,i}$ is $N_i \times N_i$, where $N_i$ is the total number of atomic orbitals for all atoms in the slice $i$. These matrices are much smaller than ${\bf H}$, and are stored in memory at the beginning of the calculation together with matrices ${\bf H}_{i,i+1}$.

Starting from the set of matrices ${\bf H}_{i,i}$ and ${\bf H}_{i,i+1}$, we implement the simplest recursive Green function algorithm aimed at getting ${\bf G}_{i,i}^r$ from which we can compute the density matrix ${\bm \rho}_i$ of slice $i$ by replacing ${\bf G}^r$ in Eq.~(\ref{eq:rho}) with ${\bf G}_{i,i}^r$. The retarded Green function ${\bf G}_{i,i}^r$ of each slices is given by:
\begin{equation}\label{eq:slabgf}
{\bf G}_{i,i}^r(E)=[E{\bf I}_{i,i}-{\bf H}_{i,i}-{\bm \Sigma}_L^{i,i}(E)-{\bm \Sigma}_R^{i,i}(E)]^{-1}.
\end{equation}
where ${\bm \Sigma}_L^{i,i}(E)$ and ${\bm \Sigma}_R^{i,i}(E)$ are the self-energies due to the rest of the device on the left and on the right, respectively, attached to slice $i$ (${\bf I}_{i,i}$ is the unit matrix of the same size as ${\bf H}_{i,i}$).

The  self-energies ${\bm \Sigma}_L^{i,i}(E)$ generated by the left side of the device attached to slide $i$ are computed through the recursive formula which starts from the self-energy of the left semi-infinite ideal electrode, ${\bm \Sigma}_L(E-eU_L) = {\bf H}_{0,1}^\dagger \cdot {\bf g}_L^r(E-eU_L) \cdot {\bf H}_{0,1}$, and proceeds through
\begin{widetext}
\begin{subequations}
\begin{eqnarray}
{\bm \Sigma}_L^{1,1}(E) & = & {\bf H}_{1,2}^\dagger \cdot [E{\bf I}_{1,1} - {\bf H}_{1,1} - {\bm \Sigma}_L(E-eU_L)]^{-1} \cdot {\bf H}_{1,2}, \label{eq:first} \\
{\bm \Sigma}_L^{2,2}(E) & = & {\bf H}_{2,3}^\dagger \cdot [E{\bf I}_{2,2} - {\bf H}_{2,2} - {\bm \Sigma}_L^{1,1}(E)]^{-1} \cdot {\bf H}_{2,3}, \\
\vdots & = & \vdots \\
{\bm \Sigma}_L^{S-1,S-1}(E) & = & {\bf H}_{S-1,S}^\dagger \cdot [E{\bf I}_{S-1,S-1} - {\bf H}_{S-1,S-1} - {\bm \Sigma}_L^{S-2,S-2}(E)]^{-1} \cdot {\bf H}_{S-1,S}.
\end{eqnarray}
\end{subequations}
\end{widetext}
Here ${\bf g}_L^r(E)$ is portion of the retarded Green function of the isolated lead connecting atoms in the edge principal layer  that is coupled to the extended central region via ${\bf H}_{0,1}$.  The same recursion starts from the right semi-infinite ideal electrode to generate the self-energies ${\bm \Sigma}_R^{i,i}(E)$, where the self-energy of the right semi-infinite ideal electrode, ${\bm \Sigma}_R(E-eU_R)={\bf H}_{S,S+1}  \cdot {\bf g}_R^r(E-eU_R) \cdot {\bf H}_{S,S+1}^\dagger$, and the Hamiltonian ${\bf H}_{S,S}$ of the first slice $S$ on the right side of the extended central region are used to construct the starting equation of the recursion analogous to Eq.~(\ref{eq:first}).

We note here that the usual simplification in \mbox{NEGF-DFT} codes is to treat the extended central region out of equilibrium while electronic structure of the ideal semi-infinite leads is computed in equilibrium, thereby ignoring the self-consistent response of the leads to the current. Although it has been pointed out~\cite{Mera2005} that this approximation can be incompatible with asymptotic charge neutrality, this is rarely taken into account. Instead of assuming that the equilibrium band structure of the leads is rigidly shifted by the bias voltage $\mp eV_{ds}/2$ applied between the macroscopic reservoirs to which they are attached, we use $\mp eU_{L,R}$ satisfying  $eV_{ds}/2 \ge eU_L > eU_R \ge -eV_{ds}/2$ as the shifts of the lead on-site energies, \mbox{${\bm \Sigma}_{L,R}(E,V_{ds})={\bm \Sigma}_{L,R}(E \mp eU_{L,R},0)$}. Here the potential $eU_{L,R}$ is adjusted after each iteration within the self-consistency loop if the total charge on slices $1$ and $S$ (obtained from ${\rm Tr}\, {\bm \rho}_1$ and ${\rm Tr}\, {\bm \rho}_S$ respectively) is found to deviate from the neutral state charge.

After the self-consistency is reached, the transmission $T(E,V_{ds})$ in Eq.~(\ref{eq:iv}) is computed from the submatrix ${\bf G}_{S,1}^r$ obtained recursively via the Dyson equation by starting from the known retarded Green function ${\bf G}_{11}^r$ (\ref{eq:slabgf}) of the first slice on the left:
\begin{equation}
{\bf G}_{i,1}^r=[E{\bf I}_{i,i} - {\bf H}_{i,i} - {\bm \Sigma}_R^{i,i}(E)]^{-1} \cdot {\bf H}_{i-1,i}^\dagger \cdot {\bf G}_{i-1,1}^r.
\end{equation}
Thus, the computational complexity of the retarded Green function evaluation is reduced from $O(N^3)$ for the full matrix inversion to $3\bar{N_i}^3(S-1) + \bar{N_i}^3 S$ operations, where $\bar{N_i}$ is the average number of atoms within the slice $i$. This means that the time required to obtain all relevant submatrices ${\bf G}_{i,i}^r$ and ${\bf G}_{S,1}^r$ for the NEGF-DFT algorithm scales linearly $O(S)$ with increasing the length of the device (i.e., the number of slices $S$).

The recursive Green function algorithm helps to resolve only one of the two key problems in the application of NEGF-DFT to large devices. The other one discussed in Sec.~\ref{sec:introduction}---numerous sharp peaks in the integrand of ${\bm \rho}_{\rm neq}$ that render real axis integration non-convergent---can be solved in principle by including the interactions~\cite{Thygesen2008,Darancet2007} within the simulated region capable of washing
out the quantum interference effects (that are, anyhow, seldom observed in devices at room temperature). For example, the inclusion of electron-electron correlation effects within the GW approximation was demonstrated~\cite{Darancet2007} to broaden or remove sharp features in the NEGFs for test systems (such as a chain of gold atoms).

In the presence of such dephasing processes, one has to resort to the full NEGF formalism~\cite{Haug2007} whose core quantities are the retarded ${\bf G}^r$ and the lesser ${\bf G}^<$ Green function describing the density of available quantum-mechanical states and how electrons occupy those quantum states, respectively. Both Green functions can be obtained from the contour-ordered Green function defined for any two time values that lie along the Kadanoff-Baym-Keldysh time contour.~\cite{Haug2007} In addition to the retarded ${\bm \Sigma}_{\rm leads}$ and the lesser ${\bm \Sigma}^<_{\rm leads}$ self-energy due to attached electrodes, the full formalism requires to compute  self-energy functionals due to many-body interactions within the sample, ${\bm \Sigma}_{\rm int}$ and ${\bm \Sigma}^<_{\rm int}$, while using conserving approximation~\cite{Thygesen2008} for their expression in terms of ${\bf G}^r$ and ${\bf G}^<$.

In the phase-coherent transport regime, ${\bm \Sigma}_{\rm int}=0$ and ${\bm \Sigma}^<_{\rm int}=0$, so that the lesser self-energy of non-interacting (i.e., mean-field or Kohn-Sham) quasiparticles can be expressed solely in terms of the retarded self-energies of the leads
\begin{equation}\label{eq:sigma}
{\bm \Sigma}^<_{\rm leads}(E)= i f(E-\mu_L){\bm \Gamma}_L(E) + i f(E-\mu_R){\bm \Gamma}_R(E).
\end{equation}
Then the Keldysh equation
\begin{equation}\label{eq:keldysh}
{\bf G}^<(E) = {\bf G}^r(E) \cdot [{\bm \Sigma}^<_{\rm leads}(E)  + {\bm \Sigma}^<_{\rm int}(E)] \cdot {\bf G}^a(E),
\end{equation}
allows to eliminate ${\bf G}^<$ as independent NEGF and express the corresponding density matrix
\begin{equation}\label{eq:rholess}
{\bm \rho}= \frac{1}{2\pi i} \int dE\, {\bf G}^<(E),
\end{equation}
using only ${\bf G}^r(E)$ and ${\bm \Sigma}_{\rm leads}(E)$, as shown explicitly by Eq.~(\ref{eq:rho}).

On the other hand, even the simplest phenomenological NEGF models of dephasing, such as ``momentum-conserving'' choice \mbox{${\bm \Sigma}_{\rm int}(E)=d {\bf G}^r(E)$} and \mbox{${\bm \Sigma}^<_{\rm int}(E)=d {\bf G}^<(E)$} ($d$ measures the dephasing strength) proposed in Ref.~\onlinecite{Golizadeh-Mojarad2007a}, require to solve Eq.~(\ref{eq:gr}) and Eq.~(\ref{eq:keldysh}) as a system of coupled matrix equations involving full size matrices in the Hilbert space of the simulated device region. For example, in the case of the dephasing model of Ref.~\onlinecite{Golizadeh-Mojarad2007a}, this means iterative solving of Eq.~(\ref{eq:gr}), with \mbox{${\bf G}^r_0 = [E - {\bf H} - {\bm \Sigma}^r_{\rm leads}(E)]^{-1}$} as the initial guess, and then using converged ${\bf G}^r$ to solve  Eq.~(\ref{eq:keldysh}) as the Sylvester equation of matrix algebra. Obviously, in this case the sparse nature of ${\bf H}$-matrix in Eq.~(\ref{eq:hmatrix}) and the corresponding recursive Green function formulas become irrelevant for reducing the time it takes to obtain all relevant NEGFs in a single step of the self-consistent loop~(\ref{eq:scloop}).

More realistic description of interactions with the extended central region is far more computationally demanding.~\cite{Thygesen2008,Darancet2007} Thus, the only route toward first-principles modeling of transport through large devices is to remain within the phase-coherent transport regime and develop algorithms that can resolve problems in the convergence of integration in ${\bm \rho}_{\rm neq}$ along the real axis, as discussed in Sec.~\ref{subsec:NonAnalytic} or by Refs.~\onlinecite{Li2007,Joon2007}.

\subsection{Quasi-Non-Equilibrium Model} \label{subsec:QuasiNonEquilibriumModel}
The DFT part of our simulation, which constructs the Hamiltonian of the central region as an input for NEGF  post-processing to obtain the device transport properties, is performed by using the \mbox{SC-EDTB} model.~\cite{Areshkin2004,Areshkin2005} This model  accounts for atomic polarization and inter-atomic charge transfer in a standard DFT-like fashion while making it possible to use a {\em minimal basis set} of four Gaussian orbitals per carbon and one orbital per hydrogen atom. The usage of such minimal basis set allows us to reduce the size
of matrices ${\bf H}_{i,i}$ and  ${\bf H}_{i,i+1}$ discussed in Sec.~\ref{sec:recursive} without loosing any of the  important aspects of {\em ab initio} input about carbon-hydrogen systems. This makes \mbox{SC-EDTB} highly advantageous when treating systems with large number of atoms.

Conceptually, \mbox{SC-EDTB} can be viewed as the pseudo-potential DFT scheme with each atom having its own atomic orbital basis set adjustable to the local atomic environment around this atom. It is a hybrid of the non-self-consistent environment-dependent tight-binding model~\cite{Tang1996} and a Gaussian-based DFT scheme.  Such adaptive behavior adequately compensates for the low precision of the minimal orthogonal basis set.  In practice, \mbox{SC-EDTB} implements the environment dependence as the parametrization of Hamiltonian matrix elements  with respect to the atomic environment, rather than the parametrization of the atomic basis set.  For example, the parameterized part of Hamiltonian matrix elements for the atom near the edge of the nanoribbon will be different from the respective matrix elements in the middle of the strip. Similarly, the in-plane Hamiltonian matrix elements for a single graphene layer will be different from the respective matrix elements in a graphene bilayer.

The \mbox{SC-EDTB} Hamiltonian matrix elements are the sums of parameterized adaptive ``TB-like'' and non-adaptive ``true DFT'' contributions.  The former mainly accounts for the covalent bonding, while the latter describes interatomic charge transfer, atomic dipole polarization, and on-site variation of exchange potential.  The extensive comparison of \mbox{SC-EDTB} with large basis set DFT calculations indicates that \mbox{SC-EDTB} produces more precise and transferable results than minimal basis set pseudo-potential DFT schemes.  At the same time, \mbox{SC-EDTB} is faster than minimal basis set pseudo-potential DFT due to: (i) faster computation of matrix elements; (ii) unit overlap matrix (i.e. orthogonal basis set); and (iii) smaller number of components used for the description of electron density (\mbox{SC-EDTB} uses ten independent components, $s^2$, $sp_x$, $sp_y$, $sp_z$, $p_x^2$, $p_y^2$, $p_z^2$, $p_xp_y$, $p_xp_z$, $p_yp_z$, to describe the electron density at a given carbon atom).  This allows us not only to capture the interatomic charge transfer, but also to account for the dipole polarization.

The compact description of electron density makes possible  efficient combination of \mbox{SC-EDTB} with convergence acceleration schemes for both equilibrium and non-equilibrium cases, as discussed in Sec.~\ref{sec:broyden}. The more detailed specification of electron density provided by standard DFT codes in local density (or some other) approximation~\cite{Fiolhais2003} will decrease the computation efficiency, but will not affect the simulation of graphene devices whose  operation is based on charge transfer at the scale larger than carbon-carbon bond length. To accommodate systems composed of tens of thousands atoms, the \mbox{SC-EDTB} part of our \mbox{NEGF-DFT} computational code also includes the possibility of multipole expansion of Coulomb potential and parallelization on distributed/shared memory systems.

Despite \mbox{$5$~\AA} cutoff radius for the orbitals used in \mbox{SC-EDTB}, the coupling Hamiltonian matrix elements between the top and the bottom graphene layers of the system depicted in Fig.~\ref{fig:SystemConfiguration} have to be masked with zeros to simulate insulating layer in between.  This causes the nonequilibrium density matrix~(\ref{eq:rho}) in the presence of the gate voltage $V_{gs}$ to evolve into two equilibrium integrals~(\ref{subeq:GIntegral})
\begin{eqnarray} \label{subeq:QuasiNonEquilibriumDensity}
\lefteqn{{\bm \rho}_{\genfrac{}{}{0pt}{}{\rm quasi}{\rm neq}} (\mu,V_{gs},T)  =  -\frac{1}{\pi} \int\limits_{-\infty}^{+\infty} dE\, {\rm Im}\, [{\bf G}^{r}(E)] f(\mu,T,E)} \nonumber \\
&&{} - \frac{1}{\pi}\int\limits_{-\infty}^{+\infty} dE\, {\rm Im}\, [{\bf G}_{\rm gate}^{r}(E)] \left\{ f(\mu+eV_{gs},T,E) \right. \nonumber  \\
&&{} - \left. f(\mu,T,E)\right\},
\end{eqnarray}
each of which is evaluated through our ``pole summation'' algorithm encoded by the formula~(\ref{subeq:GIntegralThroughResidues}). Here ${{\bf G}_{\rm gate}^{r}}$ refers to the Green function matrix Eq.~(\ref{eq:gr}) computed for the whole device, but whose all elements associated with atoms in the lower source-channel-drain layer are masked with zeros.  That is, only those matrix elements which correspond to the gate layer are allowed to be non-zero.  We assume that the self-consistency of the recursive Green function algorithm + Broyden mixing scheme (see Appendix~\ref{sec:broyden}) is reached when  $||{\bf n}^{\rm out} - {\bf n}^{\rm in}|| < 10^{-5}$, where the elements of the electron density vector ${\bf n}$ are extracted from the diagonal blocks of the corresponding ${\bm \rho}^{\rm out}_{\genfrac{}{}{0pt}{}{\rm quasi}{\rm neq}}$ and  ${\bm \rho}^{\rm out}_{\genfrac{}{}{0pt}{}{\rm quasi}{\rm neq}}$ matrices [as discussed in Sec.~\ref{sec:recursive}, only their diagonal blocks are computed from recursively generated  submatrices ${\bf G}_{i,i}^r(E)$ of the retarded Green function].

\begin{figure*}
\includegraphics[scale=0.75,angle=0]{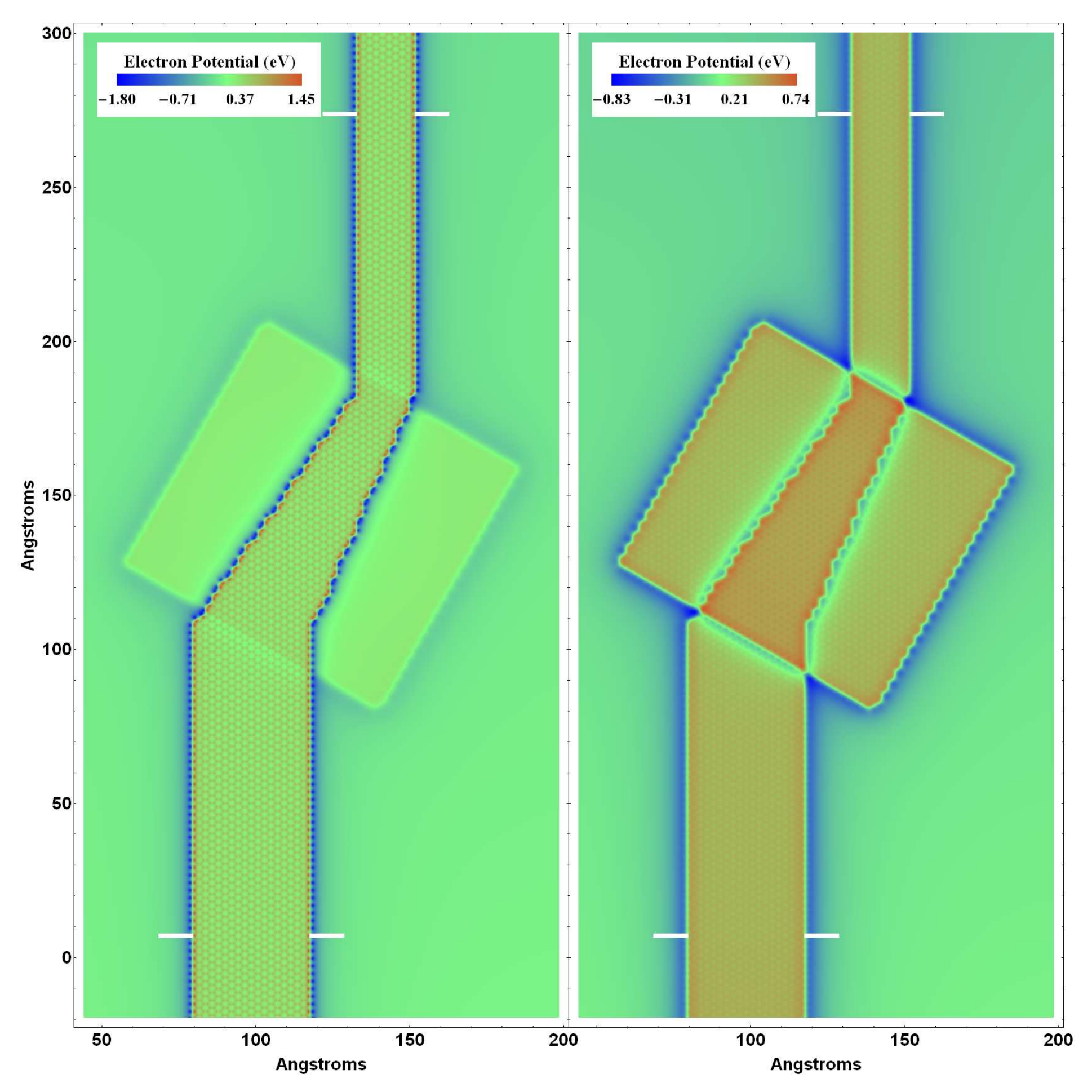}
\caption{(Color online) Contour plot of the Hartree potential for zero applied gate voltage ($V_{gs}=0$~V) in the
planes which are \mbox{0.7~\AA} (left panel) and \mbox{$0.5 \times 3.35$~\AA} (right panel) above the lower
graphene layer of the system depicted in Fig.~\ref{fig:SystemConfiguration}. White horizontal lines in the ZGNR electrode regions
mark the boundaries of the extended central region ``AGNR channel + portion of ZGNR electrodes'' composed of $\simeq 7000$ atoms (for
which the retarded Green function is evaluated to obtain electron density and electric potential through the self-consistent loop).}
\label{fig:CoulombPotentialZeroVoltage}
\end{figure*}
\begin{figure}
\includegraphics[scale=0.45,angle=0]{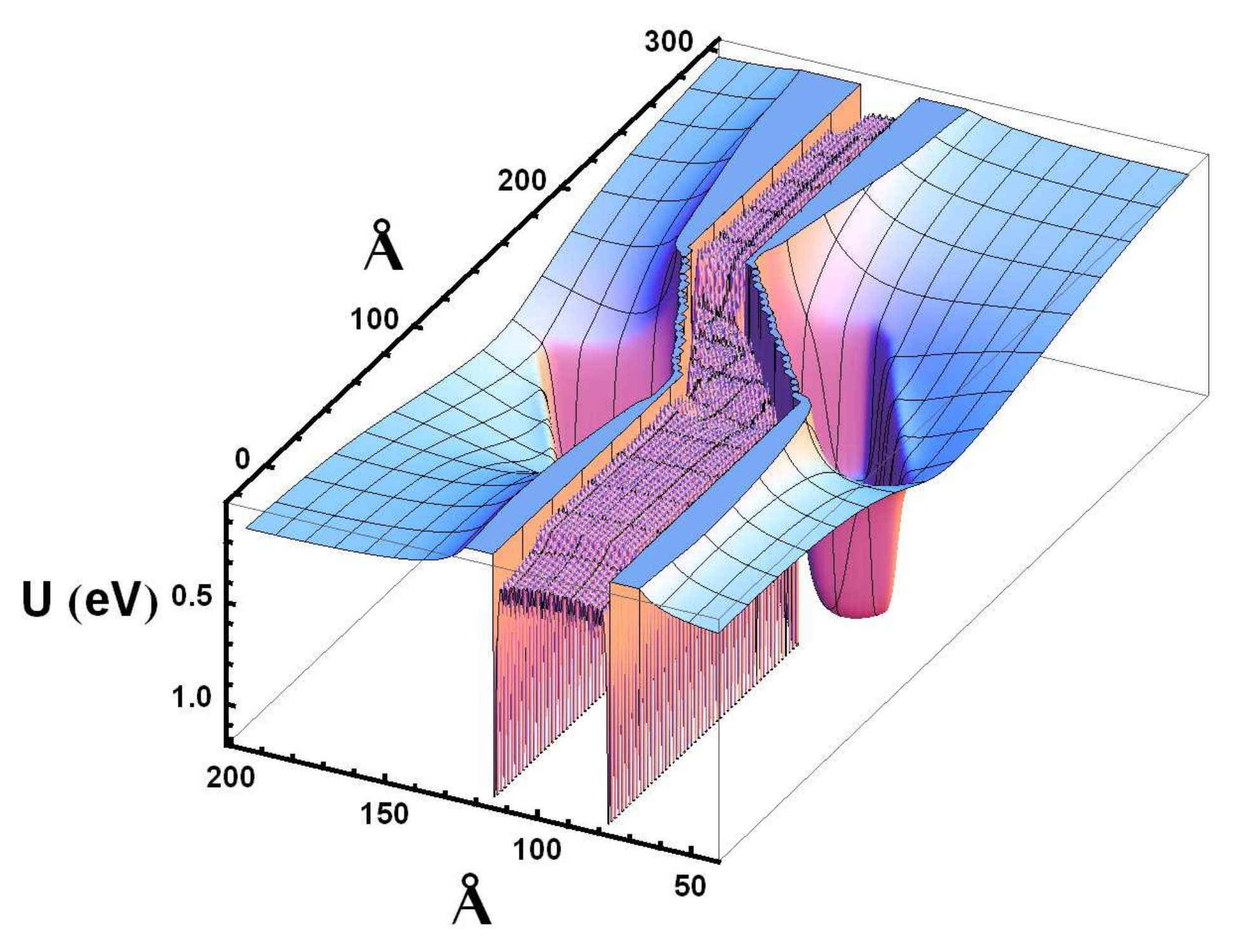}
\caption{(Color online) Contour plot of the Hartree potential in the plane \mbox{$0.2$ \AA} above the 
lower graphene layer when the applied gate voltage is \mbox{$eV_{gs}=1$~eV}.
The semiconducting region is shifted by approximately $0.35$~eV.  The potential spikes
pointing downwards correspond to the hydrogen atoms.  Positive
potential spikes associated with carbon atoms in the C-H dipole
pairs are truncated to make a clear view of the potential inside the
conducting channel. Note that the potential axis points downwards.}
\label{fig:CoulombPotential3DV1}
\end{figure}

\subsection{Results and Discussion} \label{subsec:Discussion}

We first assume zero gate voltage and plot in Fig.~\ref{fig:CoulombPotentialZeroVoltage} the self-consistent Hartree  potential~\cite{Fernandez-Rossier2007}  computed via the Poisson equation with net charge density due to charging of carbon atoms as the source term. The  potential profiles are evaluated within the planes that are parallel to two graphene layers in Fig.~\ref{fig:SystemConfiguration} and positioned in the region between them. The inhomogeneous profiles are  caused by charge transfer between hydrogen and carbon atoms.  Furthermore, it is important to emphasize that there is approximately $100$~meV difference between the Fermi levels of the wide $\mu_{\rm wide}$ and narrow $\mu_{\rm narrow}$ source and drain ZGNR electrodes, respectively,  in the bottom graphene layer of the device in Fig.~\ref{fig:SystemConfiguration}. This is caused by different ratios of carbon atoms to hydrogen atoms passivating the zigzag edges in GNRs of different widths. That is, the edge hydrogen atoms effectively dope the nanoribbon~\cite{Li2008a,Dutta2008,Biel2009} where the level of doping depends on its size and geometry. To account for this, the equilibrium Fermi level of the whole setup $\mu = (\mu_{\rm wide} + \mu_{\rm narrow})/2$  used in Eq.~(\ref{subeq:QuasiNonEquilibriumDensity}) is assumed to be the average of $\mu_{\rm wide}$ and narrow $\mu_{\rm narrow}$. Such compensation of the difference in the Fermi levels requires a small built-in electric field in our model. Room-temperature ($T = 300$ K) operation is assumed in all Figures in this Section.

\begin{figure}
\includegraphics[scale=0.55,angle=0]{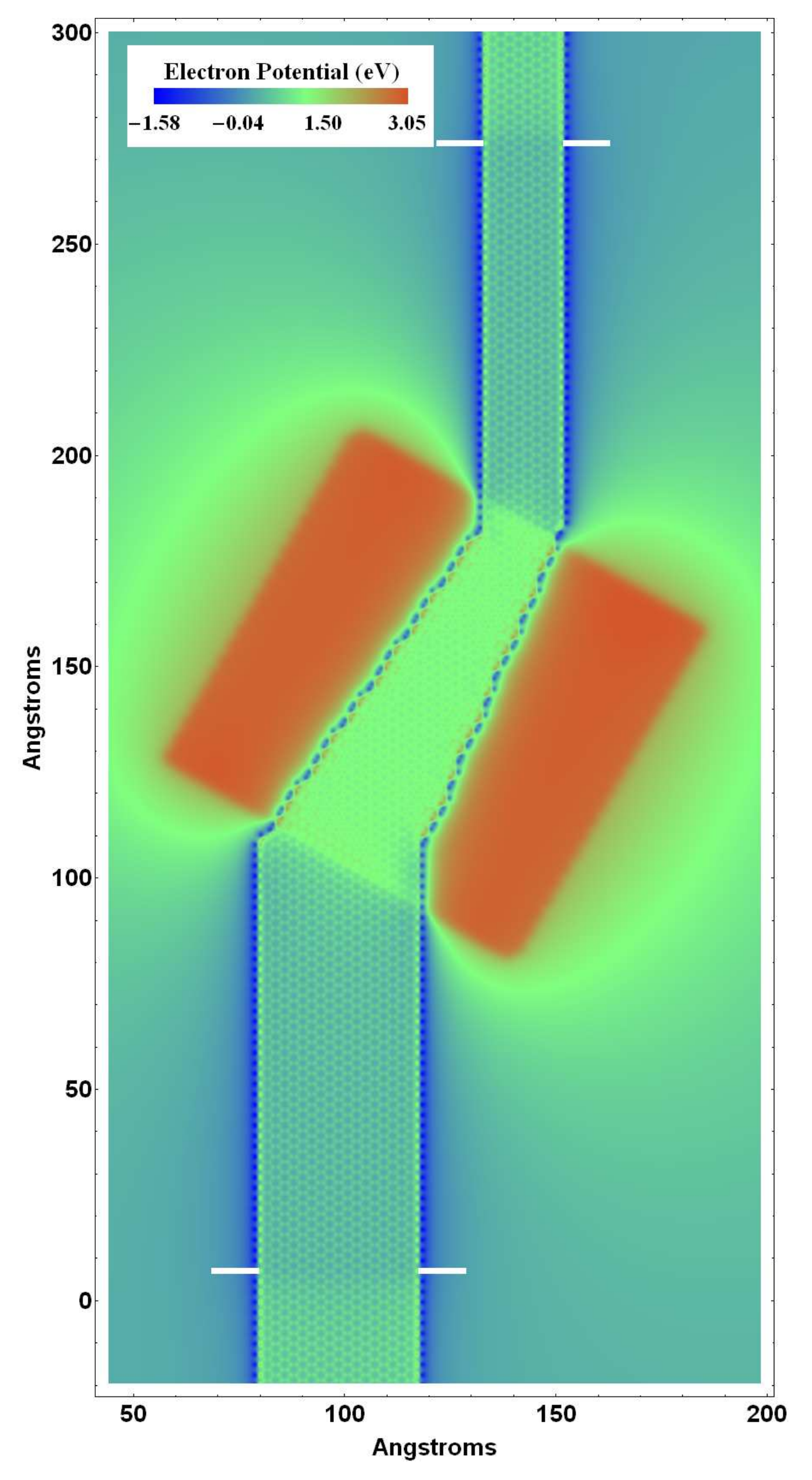}
\caption{(Color online) Contour plot of the Hartree potential for the applied gate voltage \mbox{$e V_{gs}=3$ eV} in the plane \mbox{$0.7$ \AA} above the lower graphene layer. White horizontal lines around the ZGNR electrodes mark the boundaries of the extended central region ``AGNR channel + portion of ZGNR electrodes'' composed of $\simeq 7000$ atoms.} \label{fig:CoulombPotentialThreeVolts}
\end{figure}

Then we apply voltage \mbox{$eV_{gs} = 1$ eV} to the gate electrode in Fig.~\ref{fig:CoulombPotential3DV1} and plot the full three-dimensional spatial profile of the electric potential. Further increase of the gate voltage to \mbox{$eV_{gs} = 3$ eV} leads to potential (within a geometrical plane in between two graphene layers) shown in  Fig.~\ref{fig:CoulombPotentialThreeVolts}. The self-consistent atomistic level simulation captures the potential variation in the transverse direction of the GNRs, as well as possible modifications of the band structure of GNRs  with increasing gate voltage.~\cite{Shylau2009,Fernandez-Rossier2007,Guo2007}

In both Figures, we find that the chosen portion of metallic ZGNR electrodes attached to the AGNR channel to form the ``extended central region'',~\cite{Brandbyge2002,Ke2004,Rocha2006} encompassing $\simeq 7000$ carbon and hydrogen atoms for self-consistent electron density and potential calculations, is actually not large enough (despite many ZGNR supercells included into the extended central region) to completely screen the effect of the applied electric field via the top gate electrode. This is signified by the color of the Coulomb potential at the boundaries (marked by horizontal white lines in Fig.~\ref{fig:CoulombPotentialThreeVolts}) of the ``extended central region'' not being identical to the color of the uniform potential along the semi-infinite leads. The total uncompensated charge at the boundary is approximately 0.03~$e$ for \mbox{$eV_{gs} = 1$ eV} and 0.07 $e$ for \mbox{$e V_{gs} = 3$ eV}.

\begin{figure}
\includegraphics[scale=0.5,angle=0]{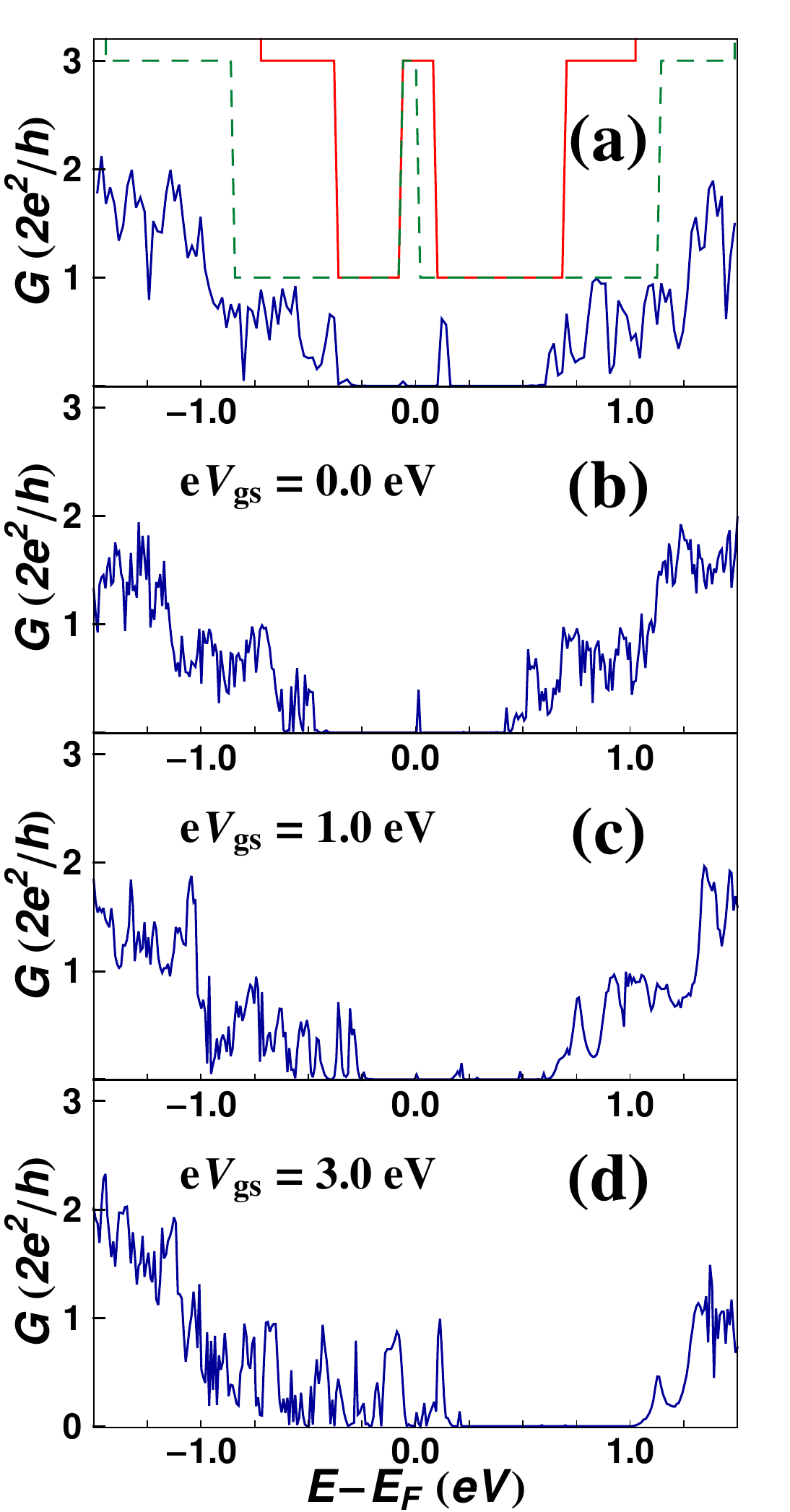}
\caption{(Color online) The non-self-consistent (a) and self-consistent (b)--(d) source-drain conductance (at linear response bias voltage $V_{ds}$) of
the nanodevice depicted in Fig.~\ref{fig:SystemConfiguration} as a function of energy. The conductances are obtained in the absence (a), (b) or presence
(c), (d) of the gate voltage $V_{gs}$, where charge redistribution is computed self-consistently in all three cases (b)--(d) [unlike in (a)].
The solid and dashed rectangular lines in panel (a) show the conductance quantization of the infinite source (wide nanoribbon, solid line) and drain (narrow nanoribbon, dashed line) ZGNR electrodes, respectively. The Fermi level in the case of unbiased gate corresponds to $E=0$.}
\label{fig:Conductance}
\end{figure}

Another feature conspicuous in Fig.~\ref{fig:CoulombPotentialThreeVolts} is that the on-site potential shift experienced by carbon atoms in the lower layer is much smaller than expected from the applied bias voltage. This unusual screening capability of insulating AGNR channel can be attributed to the presence of short segments of metallic AGNR due to either particular width of such segments (we do not relax the coordinates and edge bonds which is necessary to make all three types of AGNR insulating~\cite{Son2006a}) or doping by evanescent modes~\cite{Pomorski2004} that decay from ZGNR electrodes into AGNR channel thereby generating metal induced gap states~\cite{footnote} (localized at the ZGNR$|$AGNR interface).~\cite{Liang2008} This is also reflected in the conductance of our device---to shift the band gap of variable-width AGNR by $0.5$~eV and bring it into single channel conducting regime demands a rater large gate voltage  \mbox{$eV_{gs} \simeq 3$~eV} (when compared to \mbox{$eV_{gs} \simeq$ half-the-band-gap} required to turn uniform semiconducting AGNR into a single channel conductor~\cite{Fernandez-Rossier2007}), as shown by the source-drain conductance computed as the function of $V_{gs}$ in Figs.~\ref{fig:Conductance}(b)--(d).

The metallic behavior of ZGNR electrodes is characterized by the non-zero density of states and finite (zero temperature) conductance at the Fermi level $E_F$. We note that in simple nearest-neighbor tight-binding models~\cite{Rycerz2007} the conductance of infinite ZGNR around the charge neutral (Dirac) point $E_F=0$ is quantized $G=G_Q$ ($G_Q=2e^2/h$ is the conductance quantum for spin-degenerate transport) due to a single open conducting channel (i.e., transverse propagating mode) defined by the overlap of edge-localized wave functions.~\cite{Neto2009,Cresti2008} On the other hand, in DFT description (that can be mimicked by single $p_z$-orbital tight-binding models which include third nearest-neighbor  hopping~\cite{Cresti2008}) more complicated subband structure of ZGNR leads to three open conducting channels~\cite{Cresti2008} around $E_F=0$ and $G=3G_Q$ quantized conductance for semi-infinite source and drain ZGNR electrodes. This is confirmed in the context of our NEGF-DFT approach by Fig.~\ref{fig:Conductance}(a).

Comparing Fig.~\ref{fig:Conductance}(a) with Fig.~\ref{fig:Conductance}(b), which are both obtained at $V_{gs}=0$ V, highlights the importance of self-consistent electron density computation, even in the absence of gate voltage effects. We find a marked difference in two panels between the position of the gap region [over which the transmission function $T(E,0)$ in Eq.~(\ref{eq:transmission}) is zero]  and conductance oscillations outside of it. The conductance in Fig.~\ref{fig:Conductance}(a) was obtained without computing charge transfer effects, and could be reproduced  by popular non-self-consistent tight-binding models~\cite{Rycerz2007,Cresti2008} without resorting to full NEGF-DFT formalism.

\section{Concluding Remarks} \label{sec:conclusions}

The modeling of realistic multi-terminal graphene nanoelectronic devices requires quantum transport methods that can capture effects of its highly unusual electronic properties~\cite{Neto2009,Cresti2008} and their dependence on detailed device geometry,~\cite{Rycerz2007,Areshkin2007a} as well as  charge transfer (in equilibrium) and charge redistribution (out of equilibrium) effects on atomistic scale. While quantum transport approaches based on simple pre-defined Hamiltonians~\cite{Rycerz2007} cannot handle all of these issues, the \mbox{NEGF-DFT} framework, which generates the self-consistent Hamiltonian of the device prior to the calculation of conductance or $\mbox{\em I-V}$ characteristics, offers a proper methodology for  first-principles modeling of electron transport  involving accurate quantum-chemical description of atomic scale geometry.

However, \mbox{NEGF-DFT} simulations thus far have been limited~\cite{Stokbro2008} to rather small systems, such as short molecules connected to metallic electrodes. Here we address several obvious~\cite{Stokbro2008} and more subtle (Sec.~\ref{sec:introduction}) impediments that have to be resolved to make possible the application of \mbox{NEGF-DFT} codes to devices containing many thousand atoms: ({\em i}) computational complexity of the retarded Green function calculation, as the main time limiting part of the simulation when full Hamiltonian matrix is inverted, should scale linearly with the system size; ({\em ii}) integration of NEGFs to get the equilibrium and nonequilibrium part of the  density matrix has to be performed in a way (especially in the case of nonequilibrium contribution) which ensures convergence   despite sharp peaks (due to assumed phase-coherent transport of non-interacting quasiparticles) along the real axis whose number increases substantially in large systems; and ({\em iii}) the convergence of the self-consistent loop, which repeatedly evaluates ({\em i}) and ({\em ii}), should be accelerated with proper mixing scheme of previous iterative steps that is compatible with solution of problems in ({\em i}) and ({\em ii}).

The algorithms presented here extend the \mbox{NEGF-DFT} methodology to systems containing large number of atoms through a combination of:

\begin{list}{}{}

\item {(1)} The ``pole summation'' algorithm for the exact integration of the retarded Green function in the expression for the equilibrium part of
the density matrix offers an alternative to standard numerical contour integration by replacing the  Fermi function $f(E)$ with the analytic
function $\tilde{f}(E)$, which coincides with $f(E)$ inside the integration range along the real axis but decays exponentially in
the upper complex half-plane. Only a finite number $N_{\rm pole}$ of its poles, which can be found analytically, has non-negligible residues, so that  \mbox{${\bm \rho}_{\rm eq} = {\rm Im}\,
\sum_{j=1}^{N_{\rm pole}} \alpha_{j} {\bf G}^r(Z_{j})$}  where $\alpha_{j}$ are scalars given by simple analytical expressions in Eq.~(\ref{subeq:GIntegralThroughResidues}). The typical value of $N_{\rm pole}$ for valence electrons at room temperature is 80, and it increases with the temperature decrease with an approximate rate of 40 extra poles per order of magnitude in temperature reduction.

\item {(2)} Possible application of the  ``pole summation'' algorithm to tackle the problem of difficult-to-converge  integration of NEGFs along the real-axis (due to numerous sharp peaks in the integrand which would be impossible to locate and handle individually~\cite{Li2007,Joon2007} for devices contains large number of atoms) to obtain  ${\bm \rho}_{\rm neq}$ after its non-analytic integrand in the entire complex plane is approximated with an analytic function in the upper complex plane, so that the same type of summation can be performed as in the case of ${\bm \rho}_{\rm eq}$ integral.

\item {(3)} The recursive Green function formulas which, assuming proper geometrical decomposition of the  lattice of the device into slices of irregular shape for arbitrary nanostructure geometry, makes it possible to reduce scaling of the required computing time from $O(N^3)$ for the full Hamiltonian matrix inversion in the single iteration of the self-consistent loop to linear scaling $O(S)$ [$S$ is  the number of slices in the transport direction] of the computation of only the diagonal blocks of the retarded Green function that yield the electron density within the slice.

\end{list}

In the case of equilibrium or quasi-equilibrium (such as generated by non-zero gate voltage and zero or linear response bias voltage) situations, we additionally accelerate convergence of the self-consistent loop for the density matrix by using the modified Broyden scheme discussed in Appendix~\ref{sec:broyden}, which is compatible with the recursive Green function algorithm and  mixes input and output electron density from all previous iterations to generate  input density for the next iteration step.

We illustrate the numerical efficiency of the combination of these algorithms for NEGF part of the calculation by integrating it with the DFT code (based on the minimal  basis
set---four localized orbitals per carbon atom and one per hydrogen---tailored for carbon-hydrogen systems) to simulate gate voltage effects in \mbox{all-graphene} FET-type device. Our
simulated \mbox{ZGNR$|$variable-width-AGNR$|$ZGNR} device is composed of $\simeq 7000$  atoms and employs  AGNR of variable width (kept below $10$ nm) as a realistic semiconductor
channel accessible to present nanofabrication technology.~\cite{Li2008,Tapaszto2008,Jiao2009,Kosynkin2009} The device does not require atomic precision in controlling the width and the corresponding band gap when uniform sub-10-nm wide AGNR are used, while exploiting advantageous~\cite{Liang2008} ZGNR source and drain electrodes. We also use square-shaped gate electrode covering the channel which is made of graphene as well. The self-consistent evaluation of the electron density and Coulomb potential is required to capture inhomogeneous charge distribution and modification of the GNR band structure with increasing gate voltage.~\cite{Fernandez-Rossier2007,Shylau2009,Guo2007} This reveals that rather large gate voltage is required to shift the band gap of variable-width AGNR channel and bring this type of top-gated \mbox{GNRFET} into a window of single open transverse propagating mode with low scattering and heat dissipation.

The computation of self-consistent electron density and electrostatic potential, as the crucial aspect of NEGF-DFT approach to quantum transport modeling, is indispensable to properly take into account gate voltage effects or to ensure the gauge invariance~\cite{Christen1996} of the \mbox{\em I-V} characteristics in far from equilibrium transport.~\cite{Areshkin2009} In addition, we also demonstrate notable difference between the zero-bias transmission (i.e., linear response conductance) of non-self-consistent and self-consistent  modeling. This can be attributed to charge transfer effects between edge passivating hydrogen atoms and carbon atoms, where such edge doping also affects the position of the Fermi level of isolated GNRs of different size and geometry.

\begin{acknowledgments}
Financial support from NSF under Grant No. \mbox{ECCS 0725566} is gratefully acknowledged.
\end{acknowledgments}

\appendix

\section{Broyden mixing scheme for convergence acceleration of the self-consistent loop}\label{sec:broyden}

The recursive Green function algorithm discussed in Sec.~\ref{sec:recursive} drastically reduces the computational complexity of a single iteration step within the self-consistent  loop~(\ref{eq:scloop}). Another important ingredient of algorithms that can handle systems with large number of atoms is to combine the recursive techniques with the convergence acceleration scheme based on proper mixing of quantities found in previous steps to produce the input for the next step.

The simplest mixing scheme takes certain fraction $\varepsilon$ of the output electron density ${\bf n}^{\rm out}_m$ from the previous step $m$ and the remaining fraction $(1-\varepsilon)$ from the corresponding input ${\bf n}^{\rm in}_m$ to produce input for the next step,  ${\bf n}^{\rm in}_{m+1}=(1-\varepsilon){\bf n}^{\rm in}_m + \varepsilon {\bf n}^{\rm out}_m$. Finding the optimal value for the mixing parameter, typically $\varepsilon \sim 0.1-0.01$, depends on the nature of the system (such as, insulating vs. metallic or isolated vs. attached to semi-infinite leads). This can require few thousand iteration steps to satisfy the convergence criterion $||{\bf n}^{\rm out}_m - {\bf n}^{\rm in}_m||<10^{-5}$ we employ in our simulation.

The more sophisticated mixing schemes employ Pulay~\cite{Thygesen2008} or Broyden~\cite{Ohno2000a,Singh1986,Ihnatsenka2007} algorithms to mix several previous steps, where the quantities mixed can be the density matrix or Hamiltonian and Green functions~\cite{Thygesen2008} (which can be more efficient for open multi-terminal systems where the central region does not have a fixed number of electrons). For a small bias voltage, the self-consistency can be achieved by applying the Broyden convergence acceleration method which has two major advantages.  First,  the modified second Broyden  method~\cite{Singh1986,Ihnatsenka2007} is compatible with the recursive Green function method discussed in Sec.~\ref{sec:recursive}. Second, the Broyden method adds $O(N)$ extra operations, so that the single iteration is not slowed down.  However, the reduction of the number of iterations achieved by the Broyden method is appreciable.

The Broyden method works well when the correlation between the electron density and the potential is local, i.e., when the local potential distortion results in a local self-consistent density change.  On the other hand, in the case of non-local correlations the Broyden method performance rapidly deteriorates.  The nonequilibrium electron density in the coherent ballistic approximation constitutes the perfect example when the Broyden method fails.  The reason for this is that electron-potential correlations becomes completely non-local---the change of the potential at one contact can shut off the electron flux through the entire system and cause the system-wide electron density redistribution. Thus, in  far-from-equilibrium cases other mixing schemes have to be used.~\cite{Brandbyge2002,Areshkin2009}

In particular, the modified second Broyden method~\cite{Singh1986,Ihnatsenka2007} is compatible with the recursive Green function method discussed in Sec.~\ref{sec:recursive}, and makes it possible to reduce the number of iteration steps to the order of $\sim 10$. In this scheme, an input electron density for  iteration $m+1$ is constructed from the set of input and output densities generated in {\em all} previous iterations:
\begin{subequations}\label{eq:broyden}
\begin{eqnarray}
{\bf n}^{\rm in}_{m+1} & = & {\bf n}^{\rm in}_m - \varepsilon {\bf F}_m - \sum_{j=2}^m {\bf W}_j \cdot [{\bm \Phi}_j]^T \cdot {\bf F}_m, \\
\mbox{}  {\bf F}_m & = & {\bf n}_m^{\rm out} - {\bf n}_m^{\rm in}, \\
\mbox{} {\bf W}_i & = & -\varepsilon ({\bf F}_i - {\bf F}_{i-1}) + {\bf n}_i^{\rm in} - {\bf n}_{i-1}^{\rm in} \nonumber \\
\mbox{} && - \sum_{j=2}^{i-1} {\bf W}_j \cdot [{\bm \Phi}_j]^T \cdot ({\bf F}_i-{\bf F}_{i-1}), \\
\mbox{}  [{\bm \Phi}_i]^T & = & \frac{({\bf F}_i - {\bf F}_{i-1})^T}{({\bf F}_i - {\bf F}_{i-1})^T \cdot ({\bf F}_i - {\bf F}_{i-1})}.
\end{eqnarray}
\end{subequations}
Here ${\bf n}^{\rm in}_m$, ${\bf n}^{\rm out}_m$, ${\bf F}_m$, ${\bf W}_j$, and ${\bm \Phi}_j$ comprise a relatively small set of vectors to be stored in computer memory. The compatibility of this modified Broyden scheme with the recursive Green function algorithm of Sec.~\ref{sec:recursive} stems from the fact that only diagonal blocks of ${\bf G}^r$, required to construct vectors in Eq.~(\ref{eq:broyden}), are computed recursively without knowing the full Green function needed in some other mixing schemes.~\cite{Thygesen2008,Areshkin2009}



\begin{thebibliography}{70}
\bibitem{footnote} Although metal induced gap states do not affect the transmission for long channel devices  at linear response bias voltage,
they are expected to contribute to tunneling currents, particularly in short channel devices. In fact, they can also affect longer channel devices by
enhancing scattering processes under high source-drain bias voltage.
\end{thebibliography}

\end{document}